\documentclass[twocolumn,10pt,aps,pra,showpacs,superscriptaddress,nobalancelastpage,longbibliography,nofootinbib]{revtex4-2}

\usepackage{xcolor}
\usepackage{hyperref}  
\usepackage{nameref}   
\usepackage{graphicx}
\usepackage{amsmath,amsfonts,amssymb}
\usepackage{algorithm,algorithmicx}
\usepackage{algpseudocode}
\usepackage{hhline}
\usepackage{hyperref}
\usepackage{amsthm}
\usepackage{multirow}
\usepackage{booktabs}
\usepackage{tabularx}
\usepackage{mathtools}
\usepackage{glossaries}

\DeclarePairedDelimiter\ket{\lvert}{\rangle}

\newglossaryentry{qec}{
  name={QEC},
  description={}
}

\newglossaryentry{bb}{
  name={BB},
  description={}
}

\definecolor{C1}{RGB}{52, 89, 149}
\definecolor{C2}{RGB}{251, 77, 61}
\definecolor{C3}{RGB}{3, 206, 164}
\definecolor{C4}{RGB}{202, 21, 81}
\usepackage{hyperref}
\hypersetup{colorlinks=true, linkcolor=C2, citecolor=C2, urlcolor=C1}

\begin{document}

\title{Time-Dynamic Circuits for Fault-Tolerant Shift Automorphisms in Quantum LDPC Codes}

\author{Younghun Kim}
\email{younghunk@student.unimelb.edu.au}
\affiliation{School of Physics, The University of Melbourne, Parkville, 3010, Victoria, Australia}
\affiliation{Data61, CSIRO, Clayton, 3168, Victoria, Australia}

\author{Spiro Gicev}
\affiliation{School of Physics, The University of Melbourne, Parkville, 3010, Victoria, Australia}

\author{Martin Sevior}
\affiliation{School of Physics, The University of Melbourne, Parkville, 3010, Victoria, Australia}

\author{Muhammad Usman}
\affiliation{School of Physics, The University of Melbourne, Parkville, 3010, Victoria, Australia}
\affiliation{Data61, CSIRO, Clayton, 3168, Victoria, Australia}

\begin{abstract}
Quantum low-density parity-check (qLDPC) codes have emerged as a promising approach for realizing low-overhead logical quantum memories. Recent theoretical developments have established shift automorphisms as a fundamental building block for completing the universal set of logical gates for qLDPC codes. However, practical challenges remain because the existing SWAP-based shift automorphism yields logical error rates that are orders of magnitude higher than those for fault-tolerant idle operations. In this work, we address this issue by dynamically varying the syndrome measurement circuits to implement the shift automorphisms without reducing the circuit distance. We benchmark our approach on both twisted and untwisted weight-6 generalized toric codes, including the gross code family. Our time-dynamic circuits for shift automorphisms achieve performance comparable to the idle operations under the circuit-level noise model (SI1000). Specifically, the dynamic circuits achieve more than an order of magnitude reduction in logical error rates relative to the SWAP-based scheme for the gross code at a physical error rate of $10^{-3}$, employing the BP-OSD decoder. Our findings improve both the error resilience and the time overhead of the shift automorphisms in qLDPC codes. Furthermore, our work can lead to alternative syndrome extraction circuit designs, such as leakage removal protocols, providing a practical pathway to utilizing dynamic circuits that extend beyond surface codes towards qLDPC codes.
\end{abstract}

\maketitle

\section{INTRODUCTION}
Reliable large-scale quantum computation depends on quantum error correction (QEC) to address the fragility of physical qubits and gates \cite{shor_scheme_1995,steane_error_1996,calderbank_good_1996,knill_theory_1997,gottesman_stabilizer_1997}. Among various QEC codes, surface codes are widely considered promising for fault-tolerant quantum processor architectures due to their high threshold values and favorable compatibility with square-lattice connectivity constraints \cite{bravyi_quantum_1998,dennis_topological_2002,fowler_surface_2012,arute_quantum_2019,gong_quantum_2021}. Experimental demonstrations of small-scale surface codes have recently been reported, showing error suppression via scaling the code distance \cite{zhao_realization_2022,krinner_realizing_2022,google_quantum_ai_suppressing_2023,google_quantum_ai_and_collaborators_quantum_2025,he_experimental_2025} and the preparation of high-fidelity resource states, such as magic states essential for non-Clifford gates \cite{ye_logical_2023,kim_magic_2024,sales_rodriguez_experimental_2025,daguerre_experimental_2025,rosenfeld_magic_2025}. However, a key limitation of using surface codes for logical memory is the relatively high qubit overhead, requiring $O(d^2)$ physical qubits per logical qubit, which results in a low encoding rate. This low encoding rate poses a significant challenge to its practical fault-tolerant algorithms on current-generation quantum processors \cite{gidney_how_2021,gidney_how_2025}.
Addressing this limitation remains an important research area and alternative approaches such as densely packed configurations and hierarchical memories have been proposed \cite{gidney_how_2021,gidney_yoked_2025}.

As a promising alternative to surface codes, quantum low-density parity-check (qLDPC) codes have been developed and studied in recent literature, which promise to overcome the limitations of traditional error-correction codes \cite{gottesman_fault-tolerant_2014,breuckmann_quantum_2021}. By introducing geometric nonlocality in parity-check operators, qLDPC codes achieve higher encoding rates while maintaining error suppression comparable to surface codes \cite{bravyi_high-threshold_2024}. These features make qLDPC codes attractive for both near-term experimental demonstrations \cite{wang_demonstration_2025,andersen_small_2025} and theoretical work, including searching for better-performing variants \cite{voss_multivariate_2025,liang_generalized_2025}, developing decoders \cite{gong_toward_2024,muller_improved_2025,maurer_real-time_2025,wolanski_ambiguity_nodate,beni_tesseract_2025,hillmann_localized_2025}, and implementing logical gates through logical Pauli measurements \cite{cohen_low-overhead_2022,cross_improved_2025,he_extractors_2025}.

\begin{figure*}[t]
    \centering
    \includegraphics[width=0.95\textwidth]{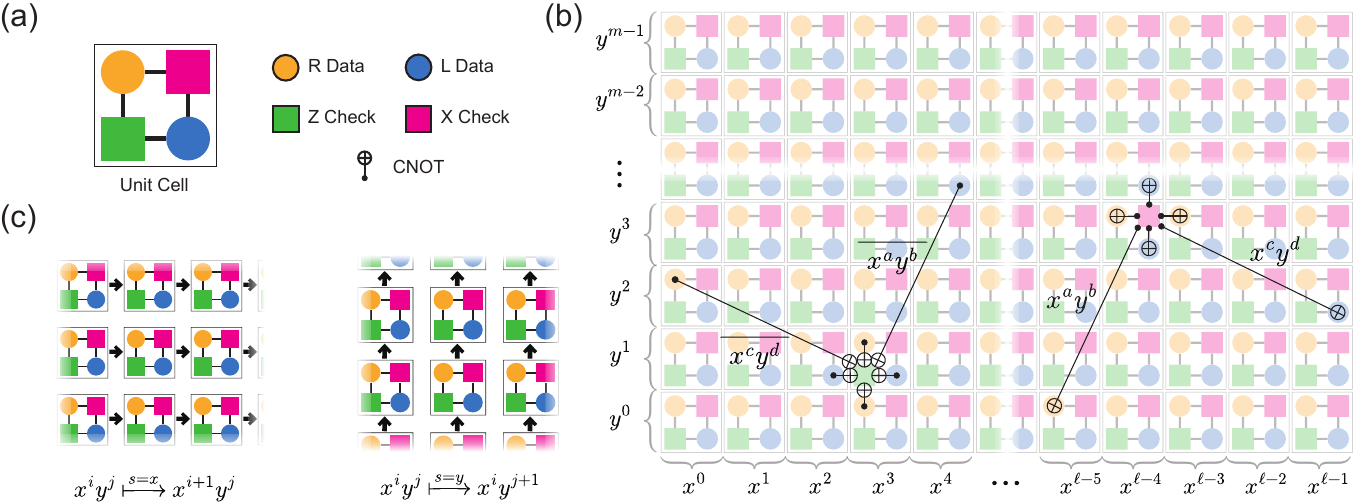}
    \caption{ \textbf{Toric layout of a weight-6 generalized toric code.} (a) Each unit cell of the code comprises two types of check operators (X and Z) and two types of data qubits (R and L), represented by distinct colors. (b) The full layout forms an $\ell\times m$ grid assembled from these unit cells, with each cell assigned coordinates defined by a polynomial representation. The CNOT gates illustrate interactions between the R and L data qubits and their corresponding X and Z check operators. The depicted layout corresponds to the gross code family, where the parameter determines the geometric nonlocality of the check operators set  $(a,b,c,d)=(-1,3,3,-1)$. (c) Each unit cell moves under the basic shift automorphisms $s=x$ and $s=y$. The diagrams illustrate the physical qubit permutations that are equivalent to the basic shift automorphisms; these operations rearrange the lattice (black arrows), effectively shifting the unit cell coordinates according to the mappings $x^i y^j \xmapsto{s=x} x^{i+1} y^j$ and $x^i y^j \xmapsto{s=y} x^i y^{j+1}$.}
    \label{fig1}
\end{figure*}

An important feature of qLDPC codes is their ability to implement logical Clifford gates and enable individual measurement of encoded logical qubits, through shift automorphisms, which can be equivalent to a logical CNOT circuit \cite{yoder_tour_2025,cross_improved_2025}. These automorphisms are conceptually achieved by translating qubits across the lattice, corresponding to global qubit permutations \cite{grassl_leveraging_2013}, illustrated in Fig.~\ref{fig1}. A shift automorphism implementation has been proposed using sequential SWAP gates acting on physically adjacent qubits \cite{bravyi_high-threshold_2024}. However, this SWAP-based approach, while fault-tolerant, suffers from significantly higher logical error rates compared to fault-tolerant idle operations. Specifically, at a physical error rate of approximately $p = 10^{-3}$ in the gross code, the logical error rate for shift automorphisms is roughly two orders of magnitude higher than that of idle operations. This performance gap widens significantly to six orders of magnitude when scaling to the two-gross code \cite{yoder_tour_2025}. Such degradation arises from the deeper circuit requirements, which involve repeated measurements, resets, and additional two-qubit gates, all of which negatively impact logical performance. Improving the time overhead and error resilience of shift automorphisms is therefore crucial for enabling efficient implementation of arbitrary logical Pauli measurements and advancing the practical use of qLDPC architectures.

Previous work has introduced time-dynamic syndrome extraction circuits in the context of the surface code, most notably through so-called \textit{walking circuits} \cite{mcewen_relaxing_2023}. These dynamic circuits have been used in previous research to periodically remove leakage \cite{eickbusch_demonstration_2025}, reposition code patches \cite{geher_error-corrected_2024,gidney_yoked_2025}, and partially deform a code patch \cite{gidney_inplace_2024}. In this work, we extend the underlying idea to qLDPC codes and show that time-dynamic syndrome extraction circuits can be utilized to implement shift automorphisms directly during syndrome measurements, thereby mitigating the high logical error rates associated with static shift automorphisms by effectively reducing the number of timesteps per full syndrome measurement to approximately two-thirds. For clarity, we refer to such operations as shift circuits throughout this paper. While our approach differs from the method proposed by Shaw et al.~\cite{shaw_lowering_2025}, which focuses on ancilla-free qLDPC codes, both techniques leverage similar circuit identities to optimize circuit-level operations.

\begin{table*}[ht]
\centering
\caption{\textbf{Parameters for the bivariant bicycle code examples.} The table lists the parameters of various bivariant bicycle (BB) codes, including the code notaion $[[n,k,d]]$, the parameter $\alpha$, the toric layout size ($\ell, m$), the polynomials $A$ and $B$ defining the X- and Z-type check operators, the code distance $d$, the circuit distance $d_{circ}$, and the modified BPT bound, the ratio $kd_{circ}^2/n$. The circuit distance is estimated using the BP-OSD decoder.}
\begin{tabular}{|c|c|c|c|c|c|c|c|c|}
\hline
$[[n,k,d]]$ & $\alpha$ & $\ell, m$ & $A$ & $B$ & $d$ & $d_{circ}$ & $kd_{circ}^2/n$ \\
\hline
$[[72,12,6]]$ & $0$ & $6,6$ & $1+x+x^{-1}y^{-3}$ & $1 + y + x^3y^{-1}$ & $6$ & $d_{circ}\le6$ & $6$\\
\hline
$[[120,8,12]]$ & $4$ & $10,6$ & $1+x+x^{-2}y$ & $1 + y + xy^2$  & $12$ & $d_{circ}\le10$ & $6.67$\\
\hline
$[[144,12,12]]$ & $0$ & $12,6$ & $1+x+x^{-1}y^{-3}$ & $1 + y + x^3y^{-1}$ & $12$ & $d_{circ}\le10$ & $8.33$\\
\hline
$[[170,16,10]]$ & $-7$ & $17,5$ & $1+x+y^{-4}$ & $1 + y + x^4$  & $10$ & $d_{circ}\le10$ & $9.41$\\
\hline
\end{tabular}
\label{tab:code_params}
\end{table*}

Building on the concept of shift circuits, our main contributions focus on time-dynamically implementing shift automorphisms in qLDPC codes. First, we design shift circuits that implement shift automorphisms within two syndrome measurement rounds, requiring only one additional layer of two-qubit gates. Furthermore, we develop a framework based on a modified coloration circuit to identify tailored syndrome cycle variants that preserve the codespace and maintain the original topological connectivity. These identified variants are then used to construct the shift circuits. Finally, we numerically evaluate the logical error rates of several \textit{bivariate bicycle} (BB) code examples listed in Table \ref{tab:code_params} \cite{kovalev_quantum_2013,lin_quantum_2024,bravyi_high-threshold_2024,liang_generalized_2025}. Our results demonstrate that the shift circuits significantly outperform the SWAP-based approach, achieving error suppression comparable to standard syndrome extraction circuits under circuit-level noise.

To validate the effectiveness of shift circuits, we conduct a detailed comparison of their logical performance against the SWAP-based approach and fault-tolerant idle operations for various BB codes. Logical error rates are assessed using Monte Carlo sampling and the failure spectrum fitting under the SI1000 noise model \cite{gidney_fault-tolerant_2021,beverland_fail_2025}. For decoding, we employ the \textit{Belief
Propagation-Ordered Statistics Decoding} (BP-OSD) decoder \cite{roffe_decoding_2020,panteleev_degenerate_2021}. Notably, at a physical error rate of $p=10^{-3}$ in the gross code, the shift circuits achieve approximately $38$-fold reduction in logical error rates relative to the SWAP-based implementations. Such results achieve logical performance within a factor of $\approx 1.7$ of fault-tolerant idle operations. These trends are consistent across twisted and untwisted weight-6 generalized toric codes studied in this work, demonstrating the robustness of the shift circuits across a range of QEC codes.

\section{METHODS}
\subsection{Generalized Toric Code}

We begin with a weight-6 generalized toric code \cite{liang_generalized_2025}, which belongs to the family of BB codes. BB codes are Calderbank-Shor-Steane (CSS)-type qLDPC codes constructed with constant-weight stabilizers for all parity-check operators and bounded qubit-degree connectivity \cite{steane_error_1996,calderbank_good_1996}.

The code consists of four types of qubits that form a unit cell on an $\ell\times m$ torus, as illustrated in Fig.~\ref{fig1}(a). Specifically, the unit cell includes two types of data qubits, denoted as $q(R,\delta)$ and $q(L,\delta)$, and two types of check qubits, $q(Z,\delta)$ and $q(X,\delta)$, which are associated with Z- and X-type check operators, respectively. Here, $\delta \in \{\, x^i y^j \mid i \in \{0, 1, \ldots, \ell - 1\},\; j \in \{0, 1, \ldots, m - 1\}\}$ represents the monomial that specifies the coordinates of the unit cell. The periodicity is defined by $m$, $\ell$, and $\alpha$ together, where $\alpha$ has the role of specifying twisting. An untwisted code corresponds to $\alpha = 0$, whereas a twisted code corresponds to $\alpha \in \mathbb{Z}$, with periodic boundary conditions as defined in Equation~\ref{eq:1}.

\begin{equation}
    \begin{aligned}
        y^m =1 \\
        x^l y^{\alpha} = 1
    \end{aligned}
\label{eq:1}
\end{equation}

A generalized toric code can be mathematically defined using two polynomials that represent the X- and Z-type Pauli check operators \cite{liang_generalized_2025}:

\begin{equation}
    \begin{aligned}
        A = A_1 +A_2 + A_3 = 1+x+x^a y^b \\
        B = B_1 +B_2 +B_3 = 1+y+x^c y^d    
    \end{aligned}
\label{eq:2}
\end{equation}

\noindent Here, $a, b, c, d \in \mathbb{Z}$ are parameters that determine the geometric nonlocality of the check operators, complementing the periodic boundary conditions. These parameters control the spatial arrangement of the check operators, which interact with data qubits across the lattice. The monomials $x^iy^j$, specified by the coordinates $\delta$ of the unit cells, define the relative positions of data qubits to check qubits. 

The X-type check operators interact with R and L data qubits through the polynomials A and B. Specifically, an X-type check qubit $q(X,\delta)$ interacts with three $R$ qubits $q(R, A_i\delta)$ via $A_i$ $(i=1,2,3)$ and three $L$ qubits $q(L, B_i\delta)$ via $B_i$ $(i=1,2,3)$. Similarly, the Z-type check operators are obtained from their linear mappings, where $x^iy^j$ is mapped to $\overline{x^iy^j} = x^{-i}y^{-j}$ term by term. This ensures that Z-type check operators interact with R and L data qubits through $\overline{B}$ and $\overline{A}$, respectively. 

\begin{equation}
    \begin{aligned}
        \overline{A} = \overline{A_1} +\overline{A_2} + \overline{A_3} = 1+x^{-1}+x^{-a} y^{-b} \\
        \overline{B} = \overline{B_1} + \overline{B_2} + \overline{B_3} = 1+y^{-1}+x^{-c} y^{-d}    
    \end{aligned}
\label{eq:3}
\end{equation}

This definition ensures that any pair of X- and Z-check operators commute, as their overlaps always involve an even number of qubits. These mutually commuting check operators form an Abelian group, known as the stabilizer group \cite{gottesman_stabilizer_1997}. The stabilizer group collectively defines the codespace as the $+1$ eigenspace of all elements. An example configuration of the parameters, $(a,b,c,d)=(-1,-3,3,-1)$, corresponding to the gross code family, is illustrated in Fig.~\ref{fig1}(b), which depicts its associated toric layout and shows how data and check qubits interact via CNOT gates.

\begin{table}[h]
    \centering
    \caption{ \textbf{Syndrome measurement cycle.} The table outlines the sequence of operations in the syndrome measurement cycle. Each step specifies the reset, CNOT operations (with the arrow indicating the control $\to$ target ordering), and measurement. The cycle consists of nine timesteps, including state preparation and measurement steps for both X- and Z-type checks. }
    \label{tab:circuit_sequence}
    \scriptsize
    \begin{tabular}{c | c }
        \toprule
        \textbf{Step} & \textbf{Operation} \\
        \midrule
        1 & $\text{Reset} : q(X,\delta) \ \& \ q(Z,\delta)$ \\
        \midrule
        2 & $\text{CNOT}_{\text{B3}} : q(R,\overline{B_3}\delta) \to q(Z,\delta)$ \\
        \midrule
        3 & $\text{CNOT}_{\text{B2}} : q(R,\overline{B_2}\delta) \to q(Z,\delta)$ \\
         & $\text{CNOT}_{\text{B1}} : q(X,\delta) \to q(L,B_1\delta)$ \\
        \midrule
        4 & $\text{CNOT}_{\text{A3}} : q(L,\overline{A_3}\delta) \to q(Z,\delta)$ \\
         & $\text{CNOT}_{\text{A1}} : q(X,\delta) \to q(R,A_1\delta)$ \\
        \midrule
        5 & $\text{CNOT}_{\text{A1}} : q(L,\overline{A_1}\delta) \to q(Z,\delta)$ \\
         & $\text{CNOT}_{\text{A3}} : q(X,\delta) \to q(R,A_3\delta)$ \\
        \midrule
        6 & $\text{CNOT}_{\text{A2}} : q(L,\overline{A_2}\delta) \to q(Z,\delta)$ \\
         & $\text{CNOT}_{\text{A2}} : q(X,\delta) \to q(R,A_2\delta)$ \\
        \midrule
        7 & $\text{CNOT}_{\text{B1}} : q(R,\overline{B_1}\delta) \to q(Z,\delta)$ \\
         & $\text{CNOT}_{\text{B3}} : q(X,\delta) \to q(L,B_3\delta)$ \\
        \midrule
        8 & $\text{CNOT}_{\text{B2}} : q(X,\delta) \to q(L,B_2\delta)$ \\
        \midrule
        9 & $\text{Meas} : q(X,\delta) \ \& \ q(Z,\delta)$ \\
        \bottomrule
    \end{tabular}
\end{table}

The syndrome extraction circuit serves as a subroutine for measuring all stabilizer check operators. Syndrome measurement data can be processed into bitstrings of \textit{detection events}, which take nontrivial values only if errors have occurred. A decoder subsequently processes the syndrome to identify and rectify errors. The syndrome extraction circuit should be parallelized while ensuring that the codespace remains unchanged. In the gross code family, as well as the weight-6 generalized toric codes, the CNOT schedules are designed with a circuit depth of seven \cite{bravyi_high-threshold_2024,yoder_tour_2025}. Each cycle of the syndrome extraction process consists of nine timesteps, including state preparation and measurement steps performed simultaneously for both X- and Z-type checks. Importantly, there are multiple variants of these cycles, distinguished by the sequence in which the two-qubit gates are executed. For example, the $[[144,12,12]]$ code has 936 distinct variants of the syndrome cycle \cite{bravyi_high-threshold_2024}. Building on the polynomial representation of generalized toric codes, one such example is detailed in Table \ref{tab:circuit_sequence}.

Table \ref{tab:code_params} highlights previously proposed BB codes that exhibit favorable parameters in relation to the Bravyi–Poulin–Terhal (BPT) bound. This bound limits the trade-off between space efficiency and code distance under geometric locality \cite{bravyi_tradeoffs_2010}. The code distance ($d$) is defined as the minimum number of physical Pauli operators that act nontrivially on the data qubits while preserving the codespace. In contrast, the circuit distance ($d_{\mathrm{circ}}$) captures the minimum weight of fault combinations that can lead to a logical error during the syndrome measurement process. In QEC, the syndrome cycle primarily determines the logical performance, as it plays a critical role in detecting errors. Accordingly, we consider the circuit distance in place of the code distance, where the bound is given by

\begin{equation} 
    \begin{aligned}
        kd_{circ}^2 \le kd^2 =O(n).
    \end{aligned}    
\end{equation}

The circuit distance is estimated using an upper bound obtained with the BP-OSD decoder, following Ref.~\cite{yoder_tour_2025}. The first and third rows of Table \ref{tab:code_params} correspond to recent examples of the gross code family. We note that the degradation of the circuit distance from 12 to 10 was also observed for the [[120,8,12]] code, similar to the gross code.

\subsection{Shift Circuits}

This section explains the design of shift circuits, which are constructed using specific variants of syndrome measurement cycles. We first provide a brief overview of shift automorphisms and then describe the process of constructing a dynamic syndrome extraction circuit, or a shift circuit, for their implementation. We note that the shift automorphisms, realized through these shift circuits, can be implemented within the connectivity limitations of standard syndrome measurements, and they are fault-tolerant. 

Shift automorphisms are logical operations that preserve the code space by globally permuting physical qubits across the lattice. These automorphisms enable the implementation of logical Clifford gates, which are essential for universal quantum computation \cite{yoder_tour_2025}, detailed in Appendix \ref{app:shifts}. A shift automorphism is defined by the monomial $s=x^ny^m$, where $n$ and $m$ are integers ($n,m \in \mathbb{Z}$). This monomial represents a global translation that shifts the coordinates of all qubits in the lattice, mapping the coordinate $\delta$ to $s\delta$, illustrated in Fig.~\ref{fig1}(c). 

The implementation of shift automorphisms involves a two-step process in which data qubits are exchanged with check qubits, alternating their positions cyclically across the lattice. These automorphisms are categorized into two types: A-type automorphisms, represented by $A_i\overline{A_j}$, and $B$-type automorphisms, represented by $B_i\overline{B_j}$, where the terms are selected from Equations \ref{eq:2} and \ref{eq:3} by choosing $i, j \in \{1, 2, 3\}$. Each automorphism is defined by a monomial that tracks the relocation of qubits, specifically detailing how check qubits are shifted to the positions of data qubits and vice versa. 

All nontrivial shift automorphisms can be constructed using two fundamental shifts, represented by $s \in \{x, y\}$. These shifts correspond to global translations across the lattice, with $s = x$ ($s = y$) belonging to the A-type (B-type) automorphism. In this work, we focus on the implementation of the A-type automorphism ($s=x$), while the construction of the B-type automorphism ($s=y$) can be achieved by appropriately modifying the shift circuits, as detailed in Appendix \ref{automorphism}.

\begin{figure}[t]
    \centering
    \includegraphics[width=0.5\textwidth]{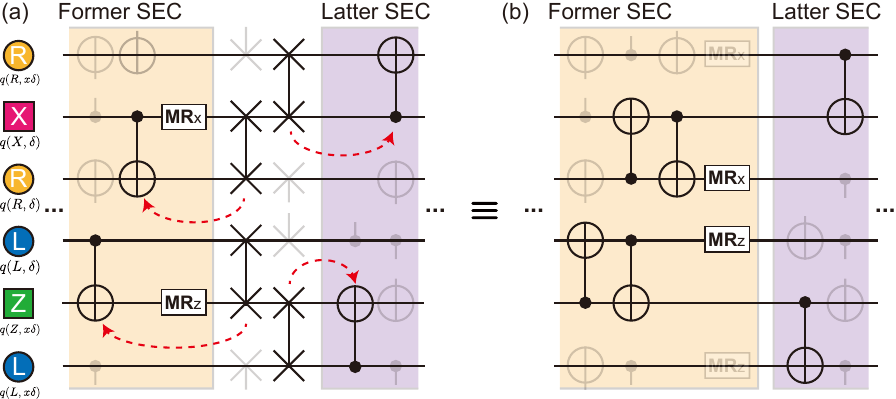}
    \caption{ \textbf{Shift circuits for implementing shift automorphisms.} (a) Two SWAP layers are inserted between the former and latter syndrome extraction cycles (SEC). The former SEC ends with CNOT gates aligned with the first SWAP layer, while the latter SEC begins with CNOT gates corresponding to the same qubit pairs as the second SWAP layer. The red dotted arrows represent where each SWAP layer is absorbed to construct a shift circuit. (b) By decomposing the SWAP gates into CNOT gates, redundant operations are canceled, effectively reducing the circuit depth. The decomposition results in a single additional layer of two-qubit gates in the former SEC and replaces syndrome qubit measurements with data qubit measurements, while simultaneously reversing the roles of control and target qubits in the initial CNOT gates of the subsequent SEC. }
    \label{fig2}
\end{figure}

For the $A$-type automorphism illustrated in Fig.~\ref{fig2}(a), each data qubit is sequentially exchanged with its adjacent check qubits in a fault-tolerant manner. This process involves alternating the positions of the qubits, acting on the L and R data qubits as follows:
\begin{equation}
\begin{array}{c@{\qquad}c}
q(R, \delta) \overset{\text{SWAP}}{\leftrightarrow} q(X, \overline{A_j}\delta) \overset{\text{SWAP}}{\leftrightarrow} q(R, A_i \overline{A_j}\delta) \\
q(L, \delta) \overset{\text{SWAP}}{\leftrightarrow} q(Z, A_i\delta) \overset{\text{SWAP}}{\leftrightarrow} q(L, A_i\overline{A_j}\delta)
\end{array}
\end{equation}
The automorphism consists of two layers of SWAP gates, followed by a syndrome cycle. As described in Ref.~\cite{yoder_tour_2025}, check qubits are typically measured and reinitialized after each SWAP gate to reduce circuit depth and define error detectors. Omitting these steps may reduce the circuit distance, which in turn can degrade the logical performance. However, in this work, we focus on constructing time-dynamic circuits and therefore omit the measurement and reinitialization steps, without observing a degradation in logical performance.  

The shift circuit incorporates two syndrome measurement cycles. The key idea relies on selecting suitable variants for the first cycle and the subsequent cycles. Initially, two layers of SWAP gates are inserted between the syndrome measurement cycles. The shift circuits are then realized by merging one layer of SWAP gates into either the preceding or succeeding syndrome cycle, as illustrated by the red dotted lines in Fig.~\ref{fig2}(a). Selecting syndrome variants with entangling gates that align with the qubit pairs targeted by the SWAP gates simplifies the circuit implementation. 

We analyze two distinct sets of syndrome cycles based on their CNOT gate configurations for the A-type local permutations. The first set begins with CNOT gates acting on the qubit pairs $(q(X, \overline{A_j}\delta),q(R, \delta))$ and $(q(L, \delta),q(Z, A_i\delta))$. The second set ends with CNOT gates acting on the pairs $(q(X, \overline{A_j}\delta),q(R, A_i \overline{A_j}\delta))$ and $(q(L, A_i \overline{A_j}\delta),q(Z, A_i\delta))$, specifying the entangling operations in the (control, target) order. These qubit-pair configurations within the syndrome cycles are key to achieving circuit-level optimization of the shift circuits.

\begin{figure}[h]
    \centering
    \includegraphics[width=0.3\textwidth]{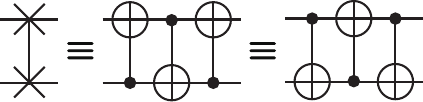}
    \caption{ \textbf{SWAP gate decomposition.} A SWAP gate can be decomposed into three CNOT gates. }
    \label{fig3}
\end{figure}

For implementation, the first and second layers of SWAP gates are first decomposed into layers of CNOT gates, as illustrated in Figure \ref{fig3}. The first SWAP layer subsequently merges with the former syndrome cycle, effectively adding one additional layer of CNOT gates and measuring data qubits instead of check qubits. Conversely, the second layer of SWAP gates flips the target and control qubits of the first gates in the latter syndrome cycle. Notably, any CNOT with a control qubit prepared in $\ket{0}$ or with a target qubit prepared in $\ket{+}$ acts as an identity operation and can therefore be omitted. These simplifications result in an equivalent circuit, as shown in Fig.~\ref{fig2}(b). The inherent variations in syndrome measurement cycles allow for multiple sequencing options in the design of shift circuits. Table \ref{tab:shift_syndrome} provides a detailed example of the complete sequence for implementing the $s=x$ shift automorphism. Crucially, the circuit distance of the shift circuit is maintained as that in idle operations.

It is worth noting that the trivial automorphism, $s=1$, represents an idle operation where qubits are not permuted, and the unit cells remain stationary. However, when using shift circuits, this is equivalent to the patch effectively stepping back and forth, and recovering its initial position every second cycle. These operations could potentially be used to periodically remove leakage errors in data qubits, as experimentally reported for the surface code family \cite{mcewen_relaxing_2023,eickbusch_demonstration_2025}.

The standard SWAP-based approach, which involves qubit measurements and reinitialization for each SWAP operation, requires four CNOT gates and one syndrome measurement circuit, resulting in a total of 15 timesteps per full syndrome measurement. On the other hand, the shift circuits, which combine two syndrome cycles and add one extra layer of CNOT gates, require 19 timesteps for the complete shift operation. This corresponds to an effective gate depth of 9.5 timesteps per syndrome, representing a significant reduction in time overhead compared to the SWAP-based method.

\begin{table}[h]
    \centering
    \caption{ \textbf{Syndrome measurement cycles of a shift circuit.} The table details the operations performed during two rounds of syndrome measurement cycles for implementing the $s=x$ shift automorphism, choosing $A_i = A_2, \overline{A_j} = \overline{A_1}$. For CNOT gates, the arrow denotes the control $\to$ target ordering.}
    \label{tab:shift_syndrome}
    \scriptsize
    \begin{tabularx}{\linewidth}{c|c|c}
        \toprule
        \textbf{Step} & \multicolumn{1}{c|}{\textbf{Round 1 Operation}} & \multicolumn{1}{c}{\textbf{Round 2 Operation}} \\
        \midrule
        
        1 & Reset: $q(X,\delta) \ \& \ q(Z,\delta)$ 
          & Reset: $q(L,\delta) \ \& \ q(R,\delta)$ \\
        \midrule
        
        2 & $\text{CNOT}_{\text{A2}}: q(L,\overline{A_2}\delta) \to q(Z,\delta)$ 
          & $\text{CNOT}_{\text{A2}}: q(Z,\delta) \to q(L,\overline{A_2}\delta)$ \\
        \midrule
        
        3 & $\text{CNOT}_{\text{A3}}: q(L,\overline{A_3}\delta) \to q(Z,\delta)$ 
          & $\text{CNOT}_{\text{A3}}: q(L,\overline{A_3}\delta) \to q(Z,\delta) $ \\
          
          & $\text{CNOT}_{\text{A1}}: q(X,\delta) \to q(R,\overline{A_1}\delta)$ 
          & $\text{CNOT}_{\text{A1}}: q(R,A_1\delta) \to q(X,\delta)$ \\
          
        \midrule
        4 & $\text{CNOT}_{\text{B3}}: q(R,\overline{B_3}\delta) \to q(Z,\delta)$ 
          & $\text{CNOT}_{\text{B3}}: q(R,\overline{B_3}\delta) \to q(Z,\delta)$ \\
          
          & $\text{CNOT}_{\text{B3}}: q(X,\delta) \to q(L,B_3\delta)$ 
          & $\text{CNOT}_{\text{B3}}: q(X,\delta) \to q(L,B_3\delta)$ \\
          
        \midrule
        5 & $\text{CNOT}_{\text{B1}}: q(R,\overline{B_1}\delta) \to q(Z,\delta)$ 
          & $\text{CNOT}_{\text{B1}}: q(R,\overline{B_1}\delta) \to q(Z,\delta)$ \\
          
          & $\text{CNOT}_{\text{B2}}: q(X,\delta) \to q(L,B_2\delta)$ 
          & $\text{CNOT}_{\text{B2}}: q(X,\delta) \to q(L,B_2\delta)$ \\
          
        \midrule
        6 & $\text{CNOT}_{\text{B2}}: q(R,\overline{B_2}\delta) \to q(Z,\delta)$ 
          & $\text{CNOT}_{\text{B2}}: q(R,\overline{B_2}\delta) \to q(Z,\delta)$ \\
          
          & $\text{CNOT}_{\text{B1}}: q(X,\delta) \to q(L,B_1\delta)$ 
          & $\text{CNOT}_{\text{B1}}: q(X,\delta) \to q(L,B_1\delta)$ \\

        \midrule
        7 & $\text{CNOT}_{\text{A1}}: q(Z,\delta) \to q(L,\overline{A_1}\delta)$ 
          & $\text{CNOT}_{\text{A1}}: q(L,\overline{A_1}\delta) \to q(Z,\delta)$ \\
          
          & $\text{CNOT}_{\text{A3}}: q(X,\delta) \to q(R,A_3\delta)$ 
          & $\text{CNOT}_{\text{A3}}: q(X,\delta) \to q(R,A_3\delta)$ \\
        \midrule
        
        8 & $\text{CNOT}_{\text{A1}}: q(L,\overline{A_1}\delta) \to q(Z,\delta)$ 
          &  \\
          
          & $\text{CNOT}_{\text{A2}}: q(R,A_2\delta) \to q(X,\delta)$ 
          & $\text{CNOT}_{\text{A2}}: q(X,\delta) \to q(R,A_2\delta)$ \\
          
        \midrule
        9 & $\text{CNOT}_{\text{A2}}: q(X,\delta) \to q(R,A_2\delta)$ 
          & $\text{Meas}: q(X,\delta) \ \& \ q(Z,\delta)$ \\
          
        \midrule
        10 & $\text{Meas}: q(L,\delta) \ \& \ q(R,\delta)$ 
           &  -  \\
        \bottomrule
    \end{tabularx}
\end{table}

Constructing shift circuits requires identifying specific variants within the syndrome measurement cycles, which are compatible with the SWAP gates. To achieve this, we utilize a modified coloration circuit to search for these variants.

\subsection{Coloration Circuit}
In this section, we detail our framework to identify the former and latter syndrome cycles using a modified version of the coloration circuit, originally introduced in Ref.~\cite{tremblay_constant-overhead_2022}. Our modifications ensure that the syndrome measurements remain deterministic, thereby guaranteeing that a syndrome cycle preserves the code space.

The coloration circuit is used to determine the scheduling of a syndrome cycle by utilizing the structure of the Tanner graph $G = (V, E)$ associated with qLDPC codes. This is a tripartite graph, with the vertex set $V$ partitioned into three subsets: one representing data qubits and the others representing X- and Z-check qubits. The edge set $E$ connects each check qubit to the data qubits involved in its corresponding check operator, whether of the X- or Z-type.

The coloration circuit assigns colors to the edges of a subgraph of the Tanner graph, such that every edge is colored. This allows the two-qubit gates corresponding to edges with the same color to be executed simultaneously. By defining a directional sequence, the coloration circuit establishes the ordering of CNOT gates, resulting in a syndrome cycle.

While the coloration circuit is broadly applicable, it is specifically designed for qLDPC code families where the check operators primarily involve up to four oriented directions: $\{e, n, w, s\}$ representing edges that connect check qubits to data qubits along specific directions: east ($e$), north ($n$), south ($s$), and west ($w$) \cite{tremblay_constant-overhead_2022}. For weight-6 generalized toric codes, we extended the coloration circuit by incorporating two additional components supported by numerical search: (1) determining the minimum circuit depth required to preserve the codespace, and (2) verifying that a given sequence of two-qubit gates produces deterministic syndromes.

It is important to note that while the same color can be assigned to directed edges for X- and Z-type check operators, some edge groups must be split to preserve the codespace. For instance, a syndrome cycle for weight-6 toric codes, particularly the gross code family, requires seven unitary rounds instead of six, which is the minimum number of colors needed to account for all directions, including non-local connections. 

\begin{table}[ht]
\centering
\caption{ \textbf{The number of possible variants for shift circuits.} The table lists the count of syndrome cycle variants that begin with an edge in the subset $\{e, n, w, s\}$ for either X or Z in the gross code. The diagonal entries indicate the primary sequences suitable for constructing shift circuits for basic shift automorphisms.}
\label{tab:syndrome_cycle_variants}
\renewcommand{\arraystretch}{1.5} 
\setlength{\tabcolsep}{12pt}      
\begin{tabular}{c|cccc}
\hline
\hline
$X \setminus Z$ & $e$ & $n$ & $w$ & $s$ \\ \hline
$e$ & 72 & 4  & 4  & 0  \\
$n$ & 4  & 72 & 0  & 4  \\
$w$ & 4  & 0  & 72 & 4  \\
$s$ & 0  & 4  & 4  & 72 \\ \hline
\hline
\end{tabular}
\end{table}

Among the variants of syndrome cycles, we will focus on syndrome cycle variants that can be oriented to begin or end with edges in the subset $\{e, n, w, s\} \subset E$ for the implementation of basic shift automorphisms for both X- and Z-checks. To identify the former syndrome cycle of the shift circuit, our framework performs a numerical search over sequences, starting with a chosen depth-7 ordering that ends with the target CNOT qubit pairs. Edges are then colored according to this sequence. Once all edges are colored, the resulting syndrome cycle is checked for deterministic syndromes. If the cycle fails to produce a valid syndrome, the process is repeated with alternative directional sequences from the CNOT schedule selection step. 

The \texttt{Stim} package is used for syndrome verification \cite{gidney_stim_2021}. Utilizing this framework, Table \ref{tab:syndrome_cycle_variants} lists the count of cycle variants that start with an edge within the subset $\{e, n, w, s\}$ for either X or Z check qubits. Notably, the number of variants terminating with these subsets is identical to the starting counts. The diagonal entries indicate the primary sequences suitable for constructing shift circuits for basic shift automorphisms. We note that the syndrome cycle variant utilized in Table \ref{tab:circuit_sequence} begins with edge $w$ and ends with edge $e$. This configuration ensures compatibility with both the former and latter syndrome circuits required to construct the basic A-type shift circuits ($s=x$).

\subsection{Noise Model and Failure Spectrum}
We consider a noise scenario that begins and ends with ideal state preparation and measurement, while the intermediate $d_{circ}$ syndrome rounds are susceptible to a circuit-level noise model, specifically the SI1000 noise model \cite{gidney_fault-tolerant_2021}. The SI1000 error model is designed to reflect the behavior of superconducting-based quantum processors. We use this model uniformly parameterized with a single error rate ($p$) to provide a practical framework for evaluating shift circuits under conditions that resemble hardware behavior when using Controlled-Z (CZ) gates as the native two-qubit gates \cite{google_quantum_ai_and_collaborators_quantum_2025,javadi-abhari_big_2025}.

To evaluate logical performance beyond the simulatable regime, at low error rates where direct sampling is computationally prohibitive within a reasonable timeframe \cite{bravyi_simulation_2013}. We employ failure-spectrum fitting that extrapolates logical error rates ($p_L$) using the following equation:

\begin{equation}
p_L = \sum_{w_0 \le w} p(\text{logical error} \,|\, w \text{ faults}) \cdot p(w \text{ faults}),
\end{equation}

\begin{figure*}[th]
    \centering
    \includegraphics[width=\textwidth]{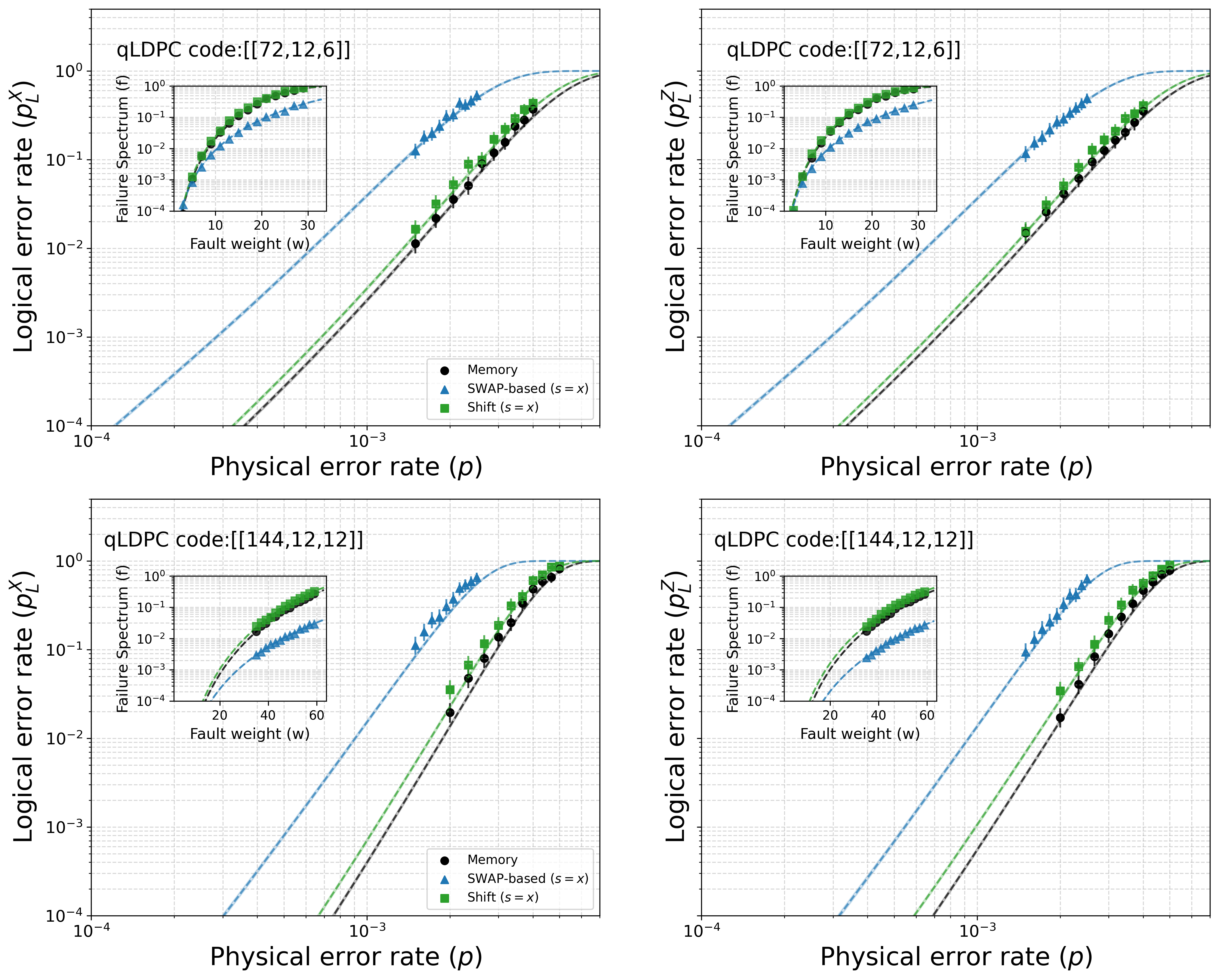}
    \caption{ \textbf{Logical error rates and failure spectrum fitting of the gross code family.} The logical performance of three logical instructions, Shift circuit (green), SWAP-based shift automorphisms (blue), and Memory (black), is evaluated for the $[[72,12,6]]$ code (top) and the gross code $[[144,12,12]]$ (bottom) using the BP-OSD decoder under the SI1000 noise model. The circuit distances for the two codes are estimated to be $d_{circ}=6$ and $d_{circ}=10$, following Ref.~\cite{yoder_tour_2025}, respectively. For each circuit, the logical error rate $p_L$ is plotted as a function of the physical error rate $p$, with data points obtained via Monte Carlo sampling and dotted lines representing the failure spectrum. Uncertainties of data points are calculated assuming binomial statistics for a $99\%$ confidence interval using \texttt{sinter} \cite{gidney_stim_2021}. Details of the fitted parameters of the failure spectrum are provided in Table \ref{tab:fit_params}. }
    \label{fig4}
\end{figure*}

\noindent which calculates the logical error rate by summing over all possible numbers of faults that could lead to a logical error, multiplying the conditional probability of a logical error given $w$ faults by the probability of $w$ faults occurring in the system. The model is based on two primary assumptions: (1) there exists a set of irreducible faults of weight $w_0$ that cause logical errors, and (2) any combination of errors leading to a logical error must include at least one of these irreducible weight $w_0$ faults \cite{beverland_fail_2025}.

To model the logical error rate, we use an ansatz based on these assumptions:

\begin{equation}
\begin{array}{c}
p_L = \displaystyle\sum_{w = w_0}^{N} f(w)\binom{N}{w} q^{w}(1 - q)^{N-w} ,
\end{array}
\end{equation}

\begin{figure*}[th]
    \centering
    \includegraphics[width=\textwidth]{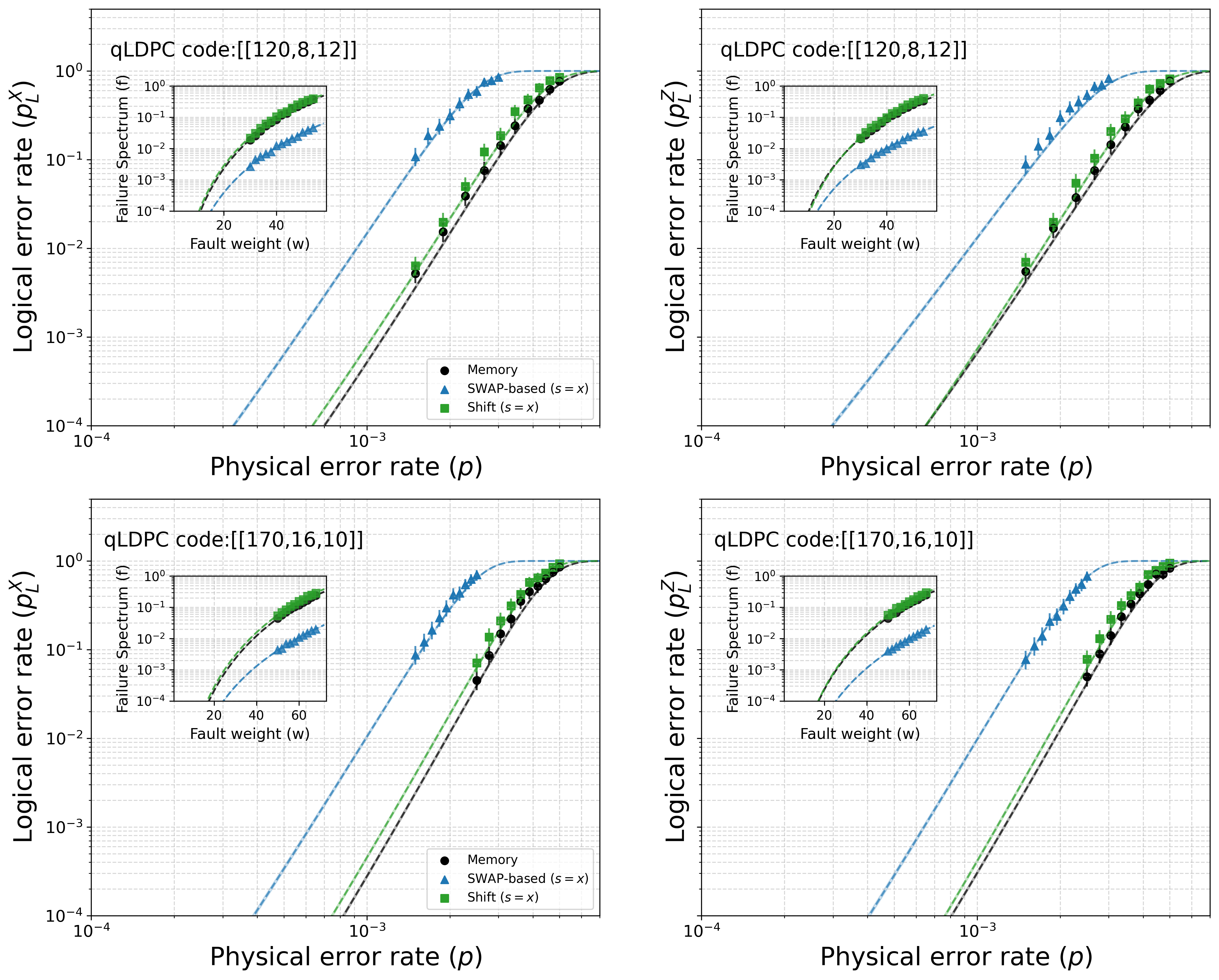}
    \caption{ \textbf{Logical error rates and failure spectrum fitting of the twisted toric codes.} The logical performance of three logical instructions, Shift circuit (green), SWAP-based shift automorphisms (blue), and Memory (black), is evaluated for the twisted toric codes, the $[[120,8,12]]$ code (top) and the $[[170,16,10]]$ code (bottom), using the BP-OSD decoder under the SI1000 noise model. The circuit distances for the two codes are estimated to be $d_{circ}=10$ for both codes, following Ref.~\cite{yoder_tour_2025}. For each circuit, the logical error rate $p_L$ is plotted as a function of the physical error rate $p$, with data points obtained via Monte Carlo sampling and dotted lines representing the failure spectrum. Uncertainties of data points are calculated assuming binomial statistics for a $99\%$ confidence interval using \texttt{sinter} \cite{gidney_stim_2021}. }
    \label{fig5}
\end{figure*}

\noindent where $w_0=\lceil d_{circ}/2\rceil$ represents the minimum weight of faults required to cause a logical error, including correction operations, $N$ is the total number of faults in the noise model, and $f(w)$ is the failure spectrum that defines the probability of logical errors given $w$ faults. To fully leverage its benefits, this approach requires fitting parameters, which we determine using Monte Carlo sampling. Specifically, we use an ansatz for $f(w)$ consisting of two fitting parameters, $f_0$ and $\gamma$. Further details on the failure spectrum can be found in Appendix \ref{failure_spec}. 

For sampling the failure spectrum, the circuit-level noise model is approximated using an alternative error rate $q = p/30$, which accounts for the combined contributions of two-qubit errors ($p/15$) and idling errors ($p/10$) in the adopted noise model. This approach assumes error sampling, where all error events are equally likely and contribute uniformly to logical errors \cite{beverland_fail_2025}.

\section{Results}
We conducted numerical simulations to evaluate the logical performance of each twisted and untwisted weight-6 toric code listed in Table \ref{tab:code_params}. Noisy circuits were constructed, and syndrome samples were generated using \texttt{Stim} \cite{gidney_stim_2021}. For decoding, the BP-OSD decoder was employed, with the decoder order set to $20$ and a maximum of $10^4$ iterations for all numerical analyses. 

The logical error rates were computed from three distinct instructions: memory experiments, SWAP-based shift automorphisms, and shift circuits. These rates were analyzed using both Monte Carlo sampling and the failure spectrum ansatz, with uncertainties for all data points calculated under the assumption of a binomial distribution. The plotted logical error rates were defined as either the logical-bit-flip error rate ($p_L^X$) or the logical-phase-flip error rate ($p_L^Z$), evaluated using state preparation and measurement in the Z-basis ($\ket{0}$) and the X-basis ($\ket{+}$). 

\begin{table}[ht]
\centering
\caption{ \textbf{Parameters for the gross code family.} Fitted parameters for the gross code family include $\text{ln}(f_0)$, $\gamma$, $w_0$, and $N$.}
\label{tab:fit_params}
\begin{tabular}{c c c c c c c}
\toprule
$[[n,k,d]]$ & Instruction & \textbf{Regime} & $\ln(f_0)$ & $\gamma$ & $w_0$ & $N$ \\

\midrule
\multirow{6}{*}{$[[72,12,6]]$}
    & \multirow{2}{*}{Memory} 
    & \textbf{Z basis} & $-8.92$ & $4.19$ & $3$ & $142560$ \\
    & & \textbf{X basis} & $-8.67$ & $4.06$ & $3$ & $142560$ \\
    \cmidrule{2-7}
    
    & \multirow{2}{*}{Shift} 
    & \textbf{Z basis} & $-8.76$ & $4.19$ & $3$ & $148716$ \\
    & & \textbf{X basis} & $-8.62$ & $4.12$ & $3$ & $148716$ \\
    \cmidrule{2-7}
    
    & \multirow{2}{*}{SWAP} 
    & \textbf{Z basis} & $-8.71$ & $3.32$ & $3$ & $439956$ \\
    & & \textbf{X basis} & $-8.81$ & $3.33$ & $3$ & $439956$ \\
    
\midrule
\multirow{6}{*}{$[[144,12,12]]$}
    & \multirow{2}{*}{Memory} 
    & \textbf{Z basis} & $-14.97$ & $5.58$ & $5$ & $475200$ \\
    & & \textbf{X basis} & $-14.17$ & $5.25$ & $5$ & $475200$ \\
    \cmidrule{2-7}
    
    & \multirow{2}{*}{Shift} 
    & \textbf{Z basis} & $-14.44$ & $5.47$ & $5$ & $495720$ \\
    & & \textbf{X basis} & $-13.47$ & $5.08$ & $5$ & $495720$ \\
    \cmidrule{2-7}
    
    & \multirow{2}{*}{SWAP} 
    & \textbf{Z basis} & $-14.5$ & $4.46$ & $5$ & $1466280$ \\
    & & \textbf{X basis} & $-14.78$ & $4.53$ & $5$ & $1466280$ \\
\bottomrule
\end{tabular}
\end{table}

Figure \ref{fig4} presents the logical error rates, $p_L^X$ and $p_L^Z$, for two codes within the gross code family, the $[[72,12,6]]$ and $[[144,12,12]]$ codes, as functions of the physical error rate $p$. The fitted parameters for the failure spectrum are detailed in Table \ref{tab:fit_params}. While the parameter $f_0$ remains relatively comparable across different logical instructions in the results, we observe a distinct difference in the parameter $\gamma$ between the SWAP-based implementation and the shift circuits. Regardless of the logical error type, the shift circuits reduce the total number of possible faults ($N$) by approximately one-third compared to the SWAP-based approach. Furthermore, the logical error rates estimated from Monte Carlo sampling and from the failure spectra are in strong agreement for both types of logical errors. 

Additionally, the shift circuits yield logical error rates comparable to those obtained from idle memory experiments across the full range of physical error rates. Specifically, the ratio $p_L^{\text{shift}}/p_L^{\text{memory}}$ is approximately $1.7$, and the protocol achieves a roughly $38$-fold reduction in the logical bit-flip error relative to the SWAP-based implementation at $p=10^{-3}$ for the $[[144,12,12]]$ gross code. This improvement is more noticeable than in the $[[72,12,6]]$ code, where the shift circuit achieves a $15$-fold reduction compared to SWAP-based results and maintains a ratio of $p_L^{\text{shift}}/p_L^{\text{memory}} \approx 1.35$. A comparable improvement is observed for logical phase errors. The failure spectrum ansatz asymptotically approaches a logical error rate of $f_0 \binom{N}{w_0} p^{w_0}$ for low $p$. By reducing $N$ to a level consistent with stationary idle operations, our framework achieves a significant reduction in the overall logical error rate, aligning with the asymptotic behavior.

Fig.~\ref{fig5} shows the logical performance of the twisted toric codes, the $[[120,8,12]]$ code and the $[[170,16,10]]$ code listed in Table \ref{tab:code_params}. The results show a behavior similar to that of the untwisted codes: shift circuits consistently align with the memory baseline, while SWAP-based implementations exhibit significantly higher logical error rates. This consistency suggests that the effectiveness of shift circuits remains robust even when applied to twisted codes.

\section{CONCLUSIONS AND OUTLOOK}

This study introduces a framework for identifying compatible syndrome cycle variants that enable shift automorphisms through the design of time-dynamic circuits. Using this framework, we conducted numerical evaluations of the logical performance of shift circuits on generalized toric codes. The findings show that these circuits achieve logical error rates comparable to fault-tolerant idle operations, significantly reducing logical error rates compared to SWAP-based implementations. Furthermore, this approach minimizes the temporal overhead by reducing the number of timesteps required for full syndrome extraction. These advancements are made possible through time-dynamic syndrome cycle designs and circuit-level optimizations that implement shift automorphisms while preserving the topological connectivity constraints and the circuit distance. In practice, any non-trivial shift automorphism can be fully realized within a logical cycle, consisting of $d_{circ}$ syndrome cycles, for both the gross and two-gross codes.

Decoding complexity is an important consideration in evaluating logical performance. Various alternative decoders, such as Relay-BP \cite{muller_improved_2025,maurer_real-time_2025}, Tesseract \cite{beni_tesseract_2025}, and BP+LSD \cite{hillmann_localized_2025}, can potentially be used for this purpose. In this work, the BP-OSD was chosen to achieve a practical balance between decoding accuracy and sampling overheads for both Monte Carlo simulations and failure-spectrum fitting. It is important to note that the logical error rates presented here were calculated before any decoder optimization, leaving room for further improvement with tailored decoding strategies. Additionally, it remains an intriguing question whether a decoder can be optimized for shift circuits along with memory, as opposed to the conventional SWAP-based implementations. A comparative analysis of logical performance across different instructions for the [[72, 12, 6]] code, utilizing the Relay-BP decoder, is presented in Appendix \ref{app:relay}. Using decoder parameters optimized exclusively for memory, including $\gamma_0 = 0.121$ and a $\gamma$ sampling range of $[-0.022, 0.182]$, we find that shift circuits maintain logical fidelity comparable to idle memory. This performance outperforms the SWAP-based implementation, even without instruction-specific tuning of the Relay-BP decoder.

Future research could focus on extending experimental demonstrations of this framework, even on a small scale, to confirm its practical feasibility beyond memory experiments. Additionally, exploring its integration with current fault-tolerant architectures, such as implementing the complete set of logical Clifford gates, would be valuable. Another promising direction involves investigating hook errors, particularly identifying syndrome cycle variants that might cause such errors and negatively impact logical performance, which was not discussed in this study. Nonetheless, our work has demonstrated that shift circuits provide avenues for significant performance improvements in shift automorphisms of generalized toric codes, thereby reducing the demands for fault-tolerant quantum computation based on these dynamic circuit architectures.

\section*{ACKNOWLEDGMENT}
YK acknowledges the support of the CSIRO Research Training Program Scholarship and the University of Melbourne Research Training Scholarship. The University of Melbourne supported the research through the establishment of the IBM Quantum Network Hub.
\\
\section*{Data Availability}
All datasets are available in the manuscript figures. Further data and source code can be made available upon reasonable request to the corresponding authors. 
\\
\section*{Author Contributions}
YK developed and simulated quantum low-density parity-check codes under the supervision of MU and MS. YK carried out all experiments and plotted figures with input from MU. YK wrote the manuscript and the Supplementary Information with input from all authors.

\bibliographystyle{quantum}


\begin{appendix}
\section{Matrix Representation of Shift Automorphisms} \label{app:shifts}

The gross code encodes twelve logical qubits across two distinct blocks, designated as prime and unprime, each containing six logical qubits. The shift automorphisms are equivalent to a sequence of logical CNOT gates that apply identical Clifford gates across both blocks. The action of these shift automorphisms can be viewed from two perspectives, depending on the chosen basis. In the computational (Z) and Hadamard (X) bases, the mappings are expressed as: $$|z\rangle \xrightarrow{s=x} |A_z z\rangle \quad \text{and} \quad |x\rangle \xrightarrow{s=x} |A_x x\rangle, $$ where $A_z$ and $A_x$ are $6 \times 6$ binary matrices that define the linear transformation of the logical qubits within each block. The transformation matrices $A_x$ and $A_z$ under the shift automorphism ($s=x$) are defined as:$$A^{gross}_z = \begin{bmatrix} 
0 & 1 & 0 & 1 & 0 & 0 \\ 
0 & 1 & 0 & 0 & 0 & 1 \\ 
0 & 0 & 1 & 1 & 0 & 0 \\ 
1 & 1 & 0 & 1 & 1 & 0 \\ 
0 & 1 & 0 & 0 & 1 & 0 \\ 
1 & 1 & 1 & 1 & 0 & 1 
\end{bmatrix}, \quad
A^{gross}_x = \begin{bmatrix} 
0 & 1 & 0 & 0 & 1 & 1 \\ 
1 & 1 & 1 & 1 & 1 & 0 \\ 
1 & 1 & 0 & 1 & 1 & 1 \\ 
0 & 1 & 1 & 1 & 1 & 1 \\ 
0 & 1 & 1 & 1 & 0 & 1 \\ 
1 & 1 & 1 & 1 & 1 & 1 
\end{bmatrix}$$

Further technical details regarding these mappings are provided in Ref.~\cite{yoder_tour_2025}.

\section{Other Shift Automorphisms} \label{automorphism}
In this session, we discuss the other type of shift automorphisms, including the B-type and non-local shift automorphisms, using shift circuits. Their \texttt{Stim} files and numerical results, as well as those for the A-type automorphism discussed in the main body, are available on GitHub\footnote{The GitHub repository will be made publicly available once prepared.}.

The other basic shift automorphism $s=y$ can be obtained by choosing $B_i=B_2$ and $\overline{B_j} = \overline{B_1}$. Likewise, in the B-type automorphism, the positions of the qubits are alternated, acting on the L and R data qubits as follows:
\begin{equation}
\begin{array}{c@{\qquad}c}
q(L, \delta) \overset{\text{SWAP}}{\leftrightarrow} q(X, \overline{B_j}\delta) \overset{\text{SWAP}}{\leftrightarrow} q(L, B_i \overline{B_j}\delta) \\
q(R, \delta) \overset{\text{SWAP}}{\leftrightarrow} q(Z, B_i\delta) \overset{\text{SWAP}}{\leftrightarrow} q(R, B_i\overline{B_j}\delta)
\end{array}
\end{equation}
Table \ref{tab:B_shift_syndrome} provides a detailed example of the complete sequence for implementing the $s=y$ shift automorphism. 

\begin{table}[h]
    \centering
    \caption{ \textbf{Syndrome measurement cycles of a shift circuit.} The table details the operations performed during two rounds of syndrome measurement cycles for implementing the $s=y$ shift automorphism, choosing $B_i = B_2, \overline{B_j} = \overline{B_1}$. For CNOT gates, the arrow denotes the control $\to$ target ordering.}
    \label{tab:B_shift_syndrome}
    \scriptsize
    \begin{tabularx}{\linewidth}{c|c|c}
        \toprule
        \textbf{Step} & \multicolumn{1}{c|}{\textbf{Round 1 Operation}} & \multicolumn{1}{c}{\textbf{Round 2 Operation}} \\
        \midrule
        
        1 & Reset: $q(X,\delta) \ \& \ q(Z,\delta)$ 
          & Reset: $q(L,\delta) \ \& \ q(R,\delta)$ \\
        \midrule
        
        2 & $\text{CNOT}_{\text{B2}}: q(R,\overline{B_2}\delta) \to q(Z,\delta)$ 
          & $\text{CNOT}_{\text{B2}}: q(Z,\delta) \to q(R,\overline{B_2}\delta)$ \\
        \midrule
        
        3 & $\text{CNOT}_{\text{B3}}: q(R,\overline{B_3}\delta) \to q(Z,\delta)$ 
          & $\text{CNOT}_{\text{B3}}: q(R,\overline{B_3}\delta) \to q(Z,\delta) $ \\
          
          & $\text{CNOT}_{\text{B1}}: q(X,\delta) \to q(L,\overline{B_1}\delta)$ 
          & $\text{CNOT}_{\text{B1}}: q(L,B_1\delta) \to q(X,\delta)$ \\
          
        \midrule
        4 & $\text{CNOT}_{\text{A2}}: q(L,\overline{A_2}\delta) \to q(Z,\delta)$ 
          & $\text{CNOT}_{\text{A2}}: q(L,\overline{A_2}\delta) \to q(Z,\delta)$ \\
          
          & $\text{CNOT}_{\text{A1}}: q(X,\delta) \to q(R,A_1\delta)$ 
          & $\text{CNOT}_{\text{A1}}: q(X,\delta) \to q(R,A_1\delta)$ \\
          
        \midrule
        5 & $\text{CNOT}_{\text{A1}}: q(L,\overline{A_1}\delta) \to q(Z,\delta)$ 
          & $\text{CNOT}_{\text{A1}}: q(L,\overline{A_1}\delta) \to q(Z,\delta)$ \\
          
          & $\text{CNOT}_{\text{A2}}: q(X,\delta) \to q(R,A_2\delta)$ 
          & $\text{CNOT}_{\text{A2}}: q(X,\delta) \to q(R,A_2\delta)$ \\
          
        \midrule
        6 & $\text{CNOT}_{\text{A3}}: q(L,\overline{A_3}\delta) \to q(Z,\delta)$ 
          & $\text{CNOT}_{\text{A3}}: q(L,\overline{A_3}\delta) \to q(Z,\delta)$ \\
          
          & $\text{CNOT}_{\text{A3}}: q(X,\delta) \to q(R,A_3\delta)$ 
          & $\text{CNOT}_{\text{A3}}: q(X,\delta) \to q(R,A_3\delta)$ \\

        \midrule
        7 & $\text{CNOT}_{\text{B1}}: q(Z,\delta) \to q(R,\overline{B_1}\delta)$ 
          & $\text{CNOT}_{\text{B1}}: q(R,\overline{B_1}\delta) \to q(Z,\delta)$ \\
          
          & $\text{CNOT}_{\text{B3}}: q(X,\delta) \to q(L,B_3\delta)$ 
          & $\text{CNOT}_{\text{B3}}: q(X,\delta) \to q(L,B_3\delta)$ \\
        \midrule
        
        8 & $\text{CNOT}_{\text{B1}}: q(R,\overline{B_1}\delta) \to q(Z,\delta)$ 
          &  \\
          
          & $\text{CNOT}_{\text{B2}}: q(L,B_2\delta) \to q(X,\delta)$ 
          & $\text{CNOT}_{\text{B2}}: q(X,\delta) \to q(L,B_2\delta)$ \\
          
        \midrule
        9 & $\text{CNOT}_{\text{B2}}: q(X,\delta) \to q(L,B_2\delta)$ 
          & $\text{Meas}: q(X,\delta) \ \& \ q(Z,\delta)$ \\
          
        \midrule
        10 & $\text{Meas}: q(L,\delta) \ \& \ q(R,\delta)$ 
           &  -  \\
        \bottomrule
    \end{tabularx}
\end{table}

We note that, in general, shift circuits can also be constructed when data qubits are permuted via non-local connections. For example, the shift automorphism $s = x^3 y^{-2}$ can be constructed by setting $B_i = B_3$ and $\bar{B}_j = \bar{B}_2$. This is achieved by identifying a variant of the syndrome measurement cycle, following the framework illustrated in the main body, that ends with $q(R,B_3\delta)$ for X checks and $q(R,\overline{B_2}\delta)$ for Z checks, and begins by interacting with $q(R,B_2\delta)$ for X checks and $q(R,\overline{B_3}\delta)$ for Z checks. Table \ref{tab:B_nonlocal_shift_syndrome} details the step-by-step sequence used to realize the shift automorphism ($s = x^3 y^{-2}$). 

Finally, we estimate the circuit distance for both the B-type shift automorphism and the non-local shift automorphism using BP+OSD. For all example codes in this work, the upper bound on the minimum number of faults causing logical errors remains the same as that for memory.

\begin{table}[h]
    \centering
    \caption{ \textbf{Syndrome measurement cycles of a shift circuit.} The table details the operations performed during two rounds of syndrome measurement cycles for implementing the $s=x^3y^{-2}$ shift automorphism, choosing $B_i = B_3, \overline{B_j} = \overline{B_2}$. For CNOT gates, the arrow denotes the control $\to$ target ordering.}
    \label{tab:B_nonlocal_shift_syndrome}
    \scriptsize
    \begin{tabularx}{\linewidth}{c|c|c}
        \toprule
        \textbf{Step} & \multicolumn{1}{c|}{\textbf{Round 1 Operation}} & \multicolumn{1}{c}{\textbf{Round 2 Operation}} \\
        \midrule
        
        1 & Reset: $q(X,\delta) \ \& \ q(Z,\delta)$ 
          & Reset: $q(L,\delta) \ \& \ q(R,\delta)$ \\
        \midrule
        
        2 & $\text{CNOT}_{\text{B3}}: q(R,\overline{B_3}\delta) \to q(Z,\delta)$ 
          & $\text{CNOT}_{\text{B3}}: q(Z,\delta) \to q(R,\overline{B_3}\delta)$ \\
        \midrule
        
        3 & $\text{CNOT}_{\text{B1}}: q(R,\overline{B_1}\delta) \to q(Z,\delta)$ 
          & $\text{CNOT}_{\text{B1}}: q(R,\overline{B_1}\delta) \to q(Z,\delta) $ \\
          
          & $\text{CNOT}_{\text{B2}}: q(X,\delta) \to q(L,\overline{B_2}\delta)$ 
          & $\text{CNOT}_{\text{B2}}: q(L,B_2\delta) \to q(X,\delta)$ \\
          
        \midrule
        4 & $\text{CNOT}_{\text{A3}}: q(L,\overline{A_3}\delta) \to q(Z,\delta)$ 
          & $\text{CNOT}_{\text{A3}}: q(L,\overline{A_3}\delta) \to q(Z,\delta)$ \\
          
          & $\text{CNOT}_{\text{A3}}: q(X,\delta) \to q(R,A_3\delta)$ 
          & $\text{CNOT}_{\text{A3}}: q(X,\delta) \to q(R,A_3\delta)$ \\
          
        \midrule
        5 & $\text{CNOT}_{\text{A2}}: q(L,\overline{A_2}\delta) \to q(Z,\delta)$ 
          & $\text{CNOT}_{\text{A2}}: q(L,\overline{A_2}\delta) \to q(Z,\delta)$ \\
          
          & $\text{CNOT}_{\text{A1}}: q(X,\delta) \to q(R,A_1\delta)$ 
          & $\text{CNOT}_{\text{A1}}: q(X,\delta) \to q(R,A_1\delta)$ \\
          
        \midrule
        6 & $\text{CNOT}_{\text{A1}}: q(L,\overline{A_1}\delta) \to q(Z,\delta)$ 
          & $\text{CNOT}_{\text{A1}}: q(L,\overline{A_1}\delta) \to q(Z,\delta)$ \\
          
          & $\text{CNOT}_{\text{A2}}: q(X,\delta) \to q(R,A_2\delta)$ 
          & $\text{CNOT}_{\text{A2}}: q(X,\delta) \to q(R,A_2\delta)$ \\

        \midrule
        7 & $\text{CNOT}_{\text{B2}}: q(Z,\delta) \to q(R,\overline{B_2}\delta)$ 
          & $\text{CNOT}_{\text{B2}}: q(R,\overline{B_2}\delta) \to q(Z,\delta)$ \\
          
          & $\text{CNOT}_{\text{B1}}: q(X,\delta) \to q(L,B_1\delta)$ 
          & $\text{CNOT}_{\text{B1}}: q(X,\delta) \to q(L,B_1\delta)$ \\
        \midrule
        
        8 & $\text{CNOT}_{\text{B2}}: q(R,\overline{B_2}\delta) \to q(Z,\delta)$ 
          &  \\
          
          & $\text{CNOT}_{\text{B3}}: q(L,B_3\delta) \to q(X,\delta)$ 
          & $\text{CNOT}_{\text{B3}}: q(X,\delta) \to q(L,B_3\delta)$ \\
          
        \midrule
        9 & $\text{CNOT}_{\text{B3}}: q(X,\delta) \to q(L,B_3\delta)$ 
          & $\text{Meas}: q(X,\delta) \ \& \ q(Z,\delta)$ \\
          
        \midrule
        10 & $\text{Meas}: q(L,\delta) \ \& \ q(R,\delta)$ 
           &  -  \\
        \bottomrule
    \end{tabularx}
\end{table}

\begin{figure*}[th]
    \centering
    \includegraphics[width=\textwidth]{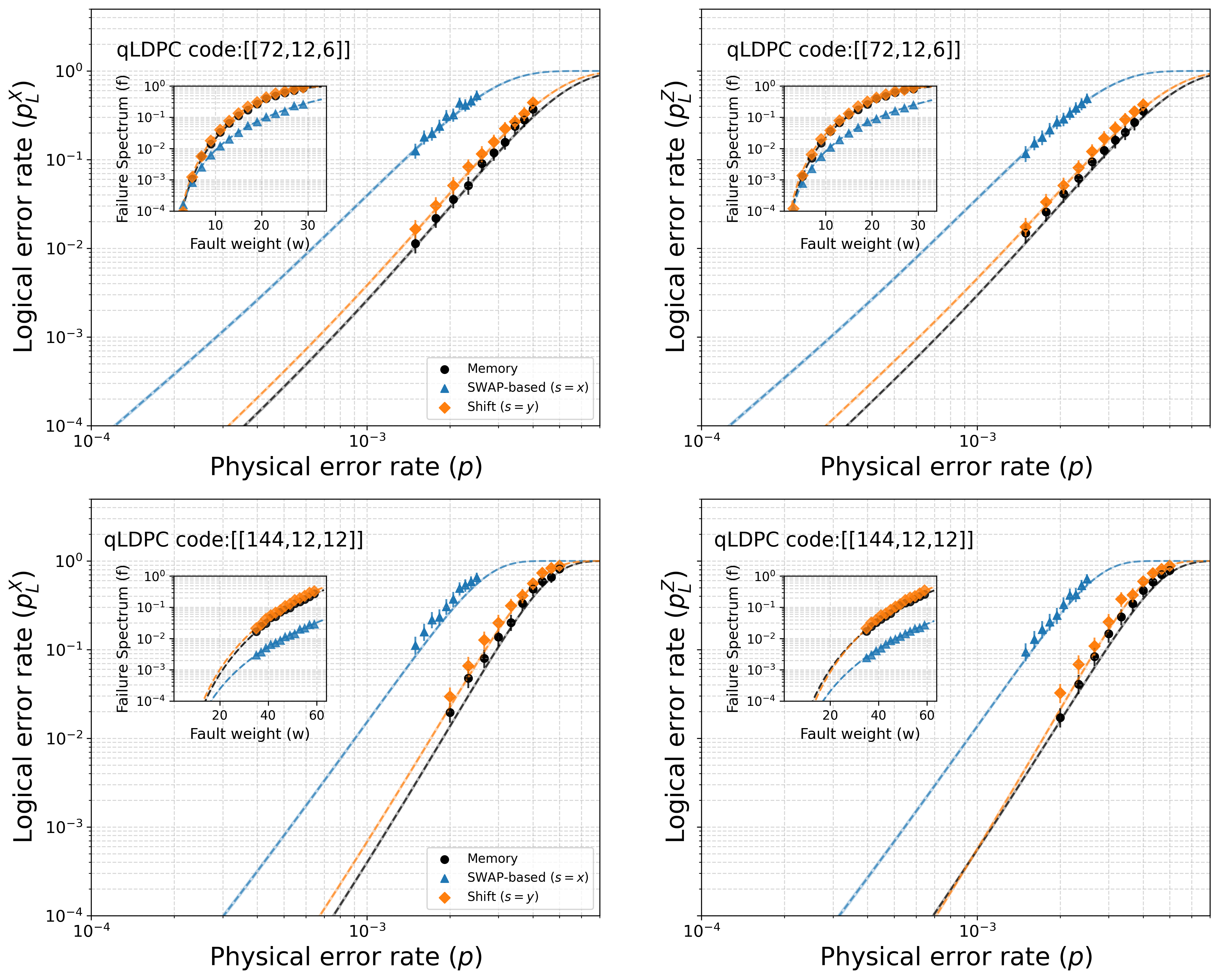}
    \caption{ \textbf{Logical error rates and failure spectrum fitting of the gross code family.} The logical performance of three logical instructions, Shift circuit (green), SWAP-based shift automorphisms (blue), and Memory (black), is evaluated for the $[[72,12,6]]$ code (top) and the gross code $[[144,12,12]]$ (bottom) using the BP-OSD decoder under the SI1000 noise model. While memory and SWAP-based results are consistent with the main body (Fig.~\ref{fig4}), the shift circuit data reflect the B-type shift automorphism ($s=y$). For each circuit, the logical error rate $p_L$ is plotted as a function of the physical error rate $p$, with data points obtained via Monte Carlo sampling and dotted lines representing the failure spectrum. }
    \label{figs1}
\end{figure*}

\begin{figure*}[th]
    \centering
    \includegraphics[width=\textwidth]{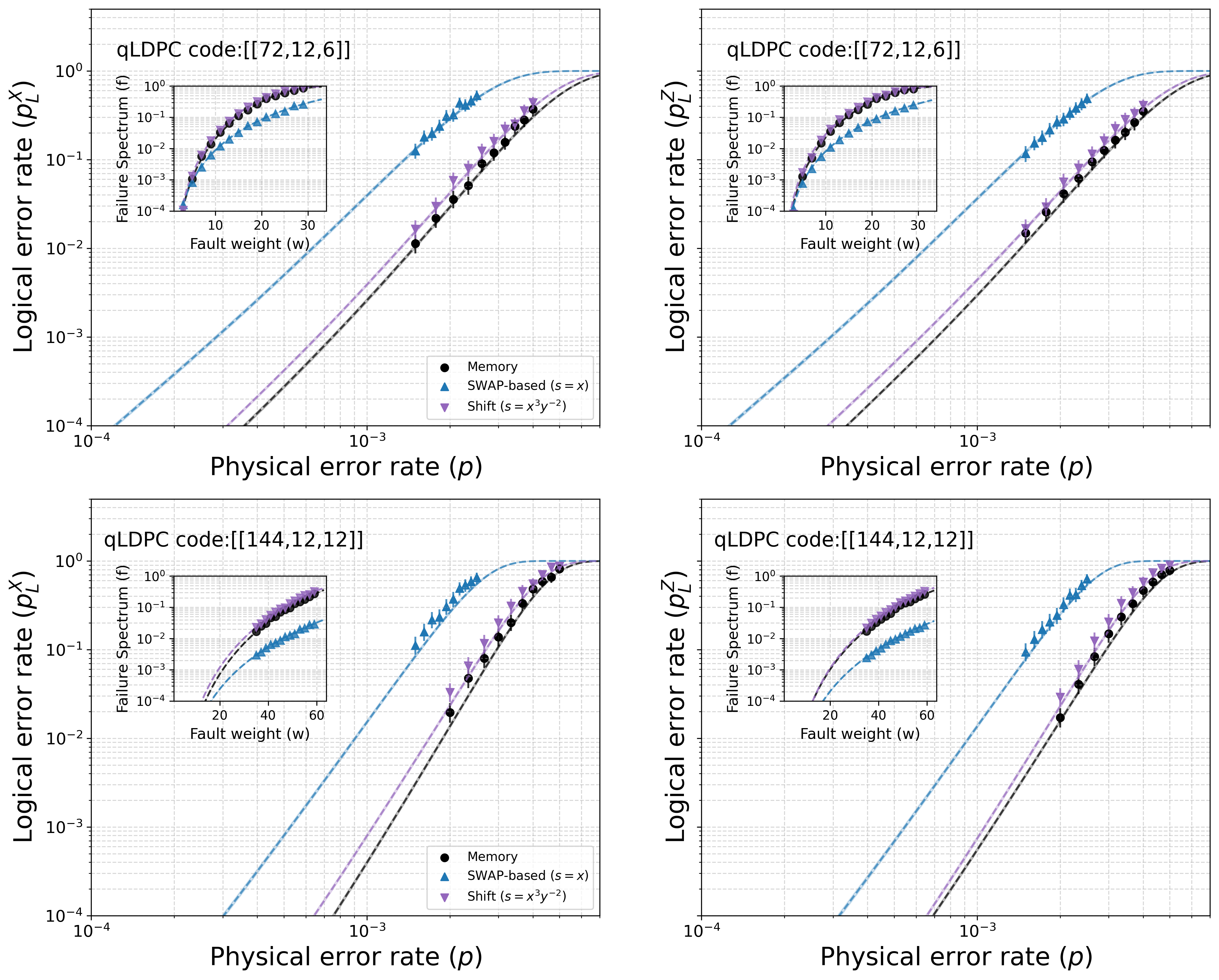}
    \caption{ \textbf{Logical error rates and failure spectrum fitting of the gross code family.} The logical performance of three logical instructions, Shift circuit (green), SWAP-based shift automorphisms (blue), and Memory (black), is evaluated for the $[[72,12,6]]$ code (top) and the gross code $[[144,12,12]]$ (bottom) using the BP-OSD decoder under the SI1000 noise model. While memory and SWAP-based results are consistent with the main body (Fig.~\ref{fig4}), the shift circuit data reflect the non-local shift automorphism ($s=x^3y^{-2}$). For each circuit, the logical error rate $p_L$ is plotted as a function of the physical error rate $p$, with data points obtained via Monte Carlo sampling and dotted lines representing the failure spectrum. }
    \label{figs2}
\end{figure*}

Figures \ref{figs1} and \ref{figs2} present the logical error rates, $p_L^X$ and $p_L^Z$, for the local B-type ($s=y$) and non-local ($s=x^3y^{-2}$) shift automorphisms as functions of the physical error rate $p$. These evaluations are performed for both the $[[72,12,6]]$ and $[[144,12,12]]$ codes. While the memory and SWAP-based implementations are consistent with the results in the main text, these figures plot the logical performance of our proposed shift circuits across different automorphisms instead of the basic A-type shift automorphism. The results demonstrate that the benefits of the shift circuit are robust and consistent, regardless of the shift automorphism types.

\section{Failure Spectrum} \label{failure_spec}
The function $f(w)$ is a well-established ansatz for estimating logical error rates across a wide range of physical error rates and QEC codes \cite{yoder_tour_2025,beverland_fail_2025}:
\begin{equation}
\begin{array}{c@{\qquad}c}
f(w)=a(1-\text{exp}(-\frac{f_0}{a}(\frac{w}{w_0})^\gamma), \ \ a=1-1/2^K
\end{array}
\end{equation}
\noindent When $w < w_0$, then $f(w) = 0$, reflecting the fact that no combination of errors can trigger a logical failure if the decoder is capable of rectifying at least $\lfloor (d-1)/2 \rfloor$ faults. The parameter $a$ is defined as the probability of a non-trivial logical error, corresponding to sampling random Pauli errors and subtracting instances of the trivial identity gate. Because our analysis accounts for either bit-flip or phase-flip logical errors among the encoded qubits, we use $K=12$. Note that we extract the parameters $f_0$ and $\gamma$ by fitting the failure spectra produced via Monte Carlo sampling.

\section{Relay-BP Decoder} \label{app:relay}
\begin{figure*}[th]
    \centering
    \includegraphics[width=\textwidth]{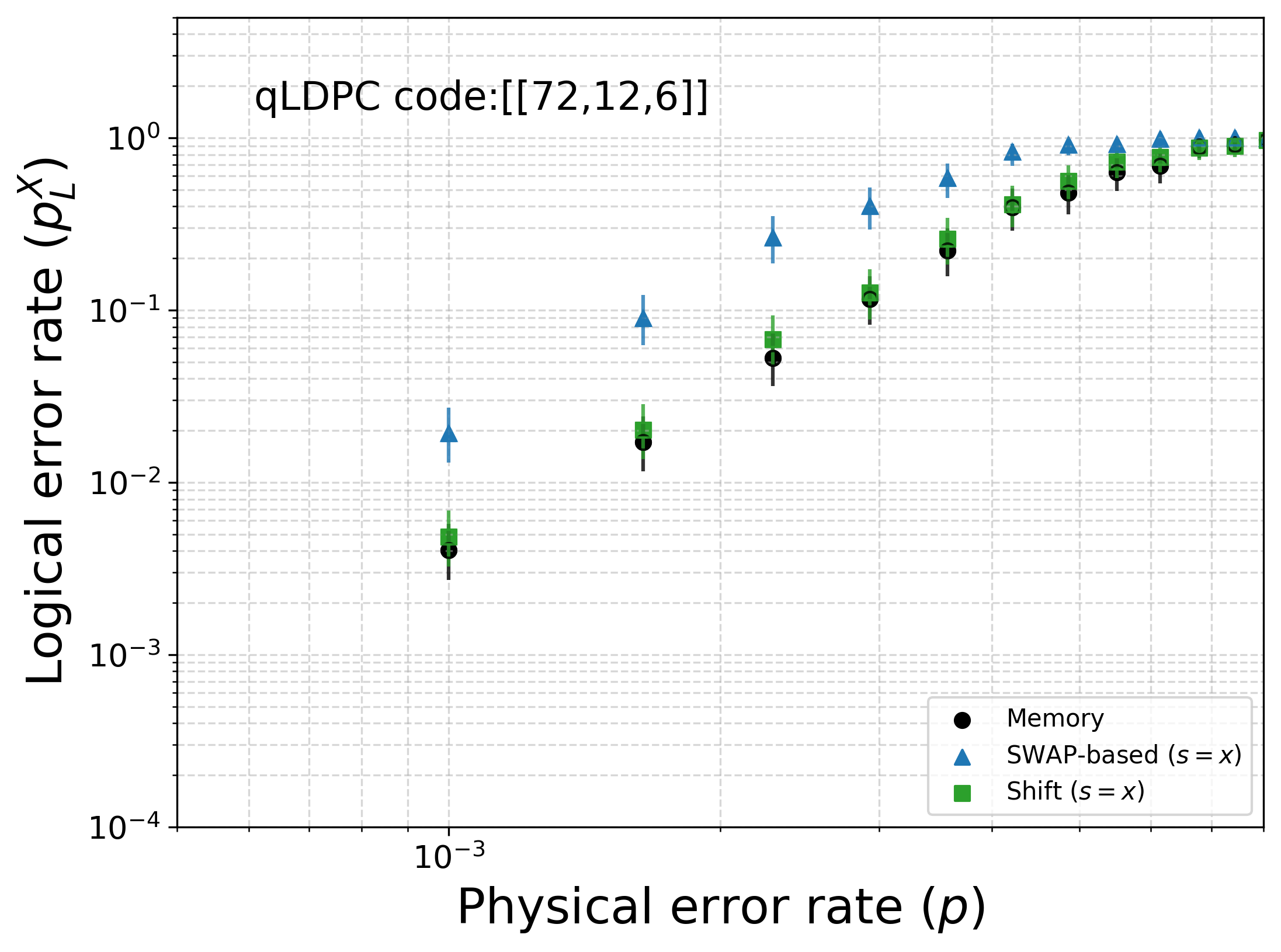}
    \caption{ \textbf{Logical error rates.} The logical performance of three logical instructions, Shift circuit (green), SWAP-based shift automorphisms (blue), and Memory (black), is evaluated for the $[[72,12,6]]$ code using the relay-BP-5 decoder under the circuit-level depolarizing noise model. The logical bit-flip error rate $p^X_L$ is plotted as a function of the physical error rate $p$, with data points obtained via Monte Carlo sampling. }
    \label{figs3}
\end{figure*}

In Fig.~\ref{figs3}, we evaluate the logical performance of three distinct instructions: memory, SWAP-based shift automorphisms, and the shift circuit, utilizing the Relay-BP decoder. Following the method in Ref.~\cite{muller_improved_2025}, we employ a circuit-level depolarizing error model to optimize the decoder for these numerical results. The parameters are fixed across all experiments to those optimized for the memory instruction: $\gamma_0 = 0.121$, $\text{pre\_iter}_1 = 601$, $\text{pre\_iter}_2 = 80$, $\text{set\_num\_iters} = 60$, and $\text{stop\_nconv} = 5$, with $\gamma$ sampled from a range $[-0.022, 0.182]$. Notably, the logical performance of the shift circuits remains comparable to that of the memory instruction, preserving the advantages over the SWAP-based implementation even without instruction-specific tuning.

\begin{thebibliography}{99}
\expandafter\ifx\csname url\endcsname\relax
  \def\url#1{\texttt{#1}}\fi
\expandafter\ifx\csname urlprefix\endcsname\relax\def\urlprefix{URL }\fi
\providecommand{\bibinfo}[2]{#2}
\providecommand{\eprint}[2][]{\url{#2}}

\bibitem{shor_scheme_1995}
\bibinfo{author}{Shor, P.~W.}
\newblock \bibinfo{title}{Scheme for reducing decoherence in quantum computer memory}.
\newblock \emph{\bibinfo{journal}{Physical Review A}} \textbf{\bibinfo{volume}{52}}, \bibinfo{pages}{R2493--R2496} (\bibinfo{year}{1995}).
\newblock \urlprefix\url{https://link.aps.org/doi/10.1103/PhysRevA.52.R2493}.

\bibitem{steane_error_1996}
\bibinfo{author}{Steane, A.~M.}
\newblock \bibinfo{title}{Error {Correcting} {Codes} in {Quantum} {Theory}}.
\newblock \emph{\bibinfo{journal}{Physical Review Letters}} \textbf{\bibinfo{volume}{77}}, \bibinfo{pages}{793--797} (\bibinfo{year}{1996}).
\newblock \urlprefix\url{https://link.aps.org/doi/10.1103/PhysRevLett.77.793}.

\bibitem{calderbank_good_1996}
\bibinfo{author}{Calderbank, A.~R.} \& \bibinfo{author}{Shor, P.~W.}
\newblock \bibinfo{title}{Good quantum error-correcting codes exist}.
\newblock \emph{\bibinfo{journal}{Physical Review A}} \textbf{\bibinfo{volume}{54}}, \bibinfo{pages}{1098--1105} (\bibinfo{year}{1996}).
\newblock \urlprefix\url{https://link.aps.org/doi/10.1103/PhysRevA.54.1098}.

\bibitem{knill_theory_1997}
\bibinfo{author}{Knill, E.} \& \bibinfo{author}{Laflamme, R.}
\newblock \bibinfo{title}{Theory of quantum error-correcting codes}.
\newblock \emph{\bibinfo{journal}{Physical Review A}} \textbf{\bibinfo{volume}{55}}, \bibinfo{pages}{900--911} (\bibinfo{year}{1997}).
\newblock \urlprefix\url{https://link.aps.org/doi/10.1103/PhysRevA.55.900}.

\bibitem{gottesman_stabilizer_1997}
\bibinfo{author}{Gottesman, D.}
\newblock \bibinfo{title}{Stabilizer {Codes} and {Quantum} {Error} {Correction}} (\bibinfo{year}{1997}).
\newblock \urlprefix\url{http://arxiv.org/abs/quant-ph/9705052}.
\newblock \bibinfo{note}{ArXiv:quant-ph/9705052}.

\bibitem{bravyi_quantum_1998}
\bibinfo{author}{Bravyi, S.~B.} \& \bibinfo{author}{Kitaev, A.~Y.}
\newblock \bibinfo{title}{Quantum codes on a lattice with boundary} (\bibinfo{year}{1998}).
\newblock \urlprefix\url{http://arxiv.org/abs/quant-ph/9811052}.
\newblock \bibinfo{note}{ArXiv:quant-ph/9811052}.

\bibitem{dennis_topological_2002}
\bibinfo{author}{Dennis, E.}, \bibinfo{author}{Kitaev, A.}, \bibinfo{author}{Landahl, A.} \& \bibinfo{author}{Preskill, J.}
\newblock \bibinfo{title}{Topological quantum memory}.
\newblock \emph{\bibinfo{journal}{Journal of Mathematical Physics}} \textbf{\bibinfo{volume}{43}}, \bibinfo{pages}{4452--4505} (\bibinfo{year}{2002}).
\newblock \urlprefix\url{http://arxiv.org/abs/quant-ph/0110143}.
\newblock \bibinfo{note}{ArXiv:quant-ph/0110143}.

\bibitem{fowler_surface_2012}
\bibinfo{author}{Fowler, A.~G.}, \bibinfo{author}{Mariantoni, M.}, \bibinfo{author}{Martinis, J.~M.} \& \bibinfo{author}{Cleland, A.~N.}
\newblock \bibinfo{title}{Surface codes: {Towards} practical large-scale quantum computation}.
\newblock \emph{\bibinfo{journal}{Physical Review A}} \textbf{\bibinfo{volume}{86}}, \bibinfo{pages}{032324} (\bibinfo{year}{2012}).
\newblock \urlprefix\url{https://link.aps.org/doi/10.1103/PhysRevA.86.032324}.

\bibitem{arute_quantum_2019}
\bibinfo{author}{Arute, F.} \emph{et~al.}
\newblock \bibinfo{title}{Quantum supremacy using a programmable superconducting processor}.
\newblock \emph{\bibinfo{journal}{Nature}} \textbf{\bibinfo{volume}{574}}, \bibinfo{pages}{505--510} (\bibinfo{year}{2019}).
\newblock \urlprefix\url{https://www.nature.com/articles/s41586-019-1666-5}.

\bibitem{gong_quantum_2021}
\bibinfo{author}{Gong, M.} \emph{et~al.}
\newblock \bibinfo{title}{Quantum walks on a programmable two-dimensional 62-qubit superconducting processor}.
\newblock \emph{\bibinfo{journal}{Science}} \textbf{\bibinfo{volume}{372}}, \bibinfo{pages}{948--952} (\bibinfo{year}{2021}).
\newblock \urlprefix\url{https://www.science.org/doi/10.1126/science.abg7812}.

\bibitem{zhao_realization_2022}
\bibinfo{author}{Zhao, Y.} \emph{et~al.}
\newblock \bibinfo{title}{Realization of an {Error}-{Correcting} {Surface} {Code} with {Superconducting} {Qubits}}.
\newblock \emph{\bibinfo{journal}{Physical Review Letters}} \textbf{\bibinfo{volume}{129}}, \bibinfo{pages}{030501} (\bibinfo{year}{2022}).
\newblock \urlprefix\url{https://link.aps.org/doi/10.1103/PhysRevLett.129.030501}.

\bibitem{krinner_realizing_2022}
\bibinfo{author}{Krinner, S.} \emph{et~al.}
\newblock \bibinfo{title}{Realizing repeated quantum error correction in a distance-three surface code}.
\newblock \emph{\bibinfo{journal}{Nature}} \textbf{\bibinfo{volume}{605}}, \bibinfo{pages}{669--674} (\bibinfo{year}{2022}).
\newblock \urlprefix\url{https://www.nature.com/articles/s41586-022-04566-8}.

\bibitem{google_quantum_ai_suppressing_2023}
\bibinfo{author}{{Google Quantum AI}} \emph{et~al.}
\newblock \bibinfo{title}{Suppressing quantum errors by scaling a surface code logical qubit}.
\newblock \emph{\bibinfo{journal}{Nature}} \textbf{\bibinfo{volume}{614}}, \bibinfo{pages}{676--681} (\bibinfo{year}{2023}).
\newblock \urlprefix\url{https://www.nature.com/articles/s41586-022-05434-1}.

\bibitem{google_quantum_ai_and_collaborators_quantum_2025}
\bibinfo{author}{{Google Quantum AI}} \emph{et~al.}
\newblock \bibinfo{title}{Quantum error correction below the surface code threshold}.
\newblock \emph{\bibinfo{journal}{Nature}} \textbf{\bibinfo{volume}{638}}, \bibinfo{pages}{920--926} (\bibinfo{year}{2025}).
\newblock \urlprefix\url{https://www.nature.com/articles/s41586-024-08449-y}.

\bibitem{he_experimental_2025}
\bibinfo{author}{He, T.} \emph{et~al.}
\newblock \bibinfo{title}{Experimental {Quantum} {Error} {Correction} below the {Surface} {Code} {Threshold} via {All}-{Microwave} {Leakage} {Suppression}}.
\newblock \emph{\bibinfo{journal}{Physical Review Letters}} \textbf{\bibinfo{volume}{135}}, \bibinfo{pages}{260601} (\bibinfo{year}{2025}).
\newblock \urlprefix\url{https://link.aps.org/doi/10.1103/rqkg-dw31}.

\bibitem{ye_logical_2023}
\bibinfo{author}{Ye, Y.} \emph{et~al.}
\newblock \bibinfo{title}{Logical {Magic} {State} {Preparation} with {Fidelity} beyond the {Distillation} {Threshold} on a {Superconducting} {Quantum} {Processor}}.
\newblock \emph{\bibinfo{journal}{Physical Review Letters}} \textbf{\bibinfo{volume}{131}}, \bibinfo{pages}{210603} (\bibinfo{year}{2023}).
\newblock \urlprefix\url{https://link.aps.org/doi/10.1103/PhysRevLett.131.210603}.

\bibitem{kim_magic_2024}
\bibinfo{author}{Kim, Y.}, \bibinfo{author}{Sevior, M.} \& \bibinfo{author}{Usman, M.}
\newblock \bibinfo{title}{Magic {State} {Injection} on {IBM} {Quantum} {Processors} {Above} the {Distillation} {Threshold}} (\bibinfo{year}{2024}).
\newblock \urlprefix\url{http://arxiv.org/abs/2412.01446}.
\newblock \bibinfo{note}{ArXiv:2412.01446 [quant-ph]}.

\bibitem{sales_rodriguez_experimental_2025}
\bibinfo{author}{Sales~Rodriguez, P.} \emph{et~al.}
\newblock \bibinfo{title}{Experimental demonstration of logical magic state distillation}.
\newblock \emph{\bibinfo{journal}{Nature}} \textbf{\bibinfo{volume}{645}}, \bibinfo{pages}{620--625} (\bibinfo{year}{2025}).
\newblock \urlprefix\url{https://www.nature.com/articles/s41586-025-09367-3}.

\bibitem{daguerre_experimental_2025}
\bibinfo{author}{Daguerre, L.}, \bibinfo{author}{Blume-Kohout, R.}, \bibinfo{author}{Brown, N.~C.}, \bibinfo{author}{Hayes, D.} \& \bibinfo{author}{Kim, I.~H.}
\newblock \bibinfo{title}{Experimental {Demonstration} of {High}-{Fidelity} {Logical} {Magic} {States} from {Code} {Switching}}.
\newblock \emph{\bibinfo{journal}{Physical Review X}} \textbf{\bibinfo{volume}{15}}, \bibinfo{pages}{041008} (\bibinfo{year}{2025}).
\newblock \urlprefix\url{https://link.aps.org/doi/10.1103/dck4-x9c2}.

\bibitem{rosenfeld_magic_2025}
\bibinfo{author}{Rosenfeld, E.} \emph{et~al.}
\newblock \bibinfo{title}{Magic state cultivation on a superconducting quantum processor} (\bibinfo{year}{2025}).
\newblock \urlprefix\url{http://arxiv.org/abs/2512.13908}.
\newblock \bibinfo{note}{ArXiv:2512.13908 [quant-ph]}.

\bibitem{gidney_how_2021}
\bibinfo{author}{Gidney, C.} \& \bibinfo{author}{Ekerå, M.}
\newblock \bibinfo{title}{How to factor 2048 bit {RSA} integers in 8 hours using 20 million noisy qubits}.
\newblock \emph{\bibinfo{journal}{Quantum}} \textbf{\bibinfo{volume}{5}}, \bibinfo{pages}{433} (\bibinfo{year}{2021}).
\newblock \urlprefix\url{http://arxiv.org/abs/1905.09749}.
\newblock \bibinfo{note}{ArXiv:1905.09749 [quant-ph]}.

\bibitem{gidney_how_2025}
\bibinfo{author}{Gidney, C.}
\newblock \bibinfo{title}{How to factor 2048 bit {RSA} integers with less than a million noisy qubits} (\bibinfo{year}{2025}).
\newblock \urlprefix\url{http://arxiv.org/abs/2505.15917}.
\newblock \bibinfo{note}{ArXiv:2505.15917 [quant-ph]}.

\bibitem{gidney_yoked_2025}
\bibinfo{author}{Gidney, C.}, \bibinfo{author}{Newman, M.}, \bibinfo{author}{Brooks, P.} \& \bibinfo{author}{Jones, C.}
\newblock \bibinfo{title}{Yoked surface codes}.
\newblock \emph{\bibinfo{journal}{Nature Communications}} \textbf{\bibinfo{volume}{16}}, \bibinfo{pages}{4498} (\bibinfo{year}{2025}).
\newblock \urlprefix\url{https://www.nature.com/articles/s41467-025-59714-1}.

\bibitem{gottesman_fault-tolerant_2014}
\bibinfo{author}{Gottesman, D.}
\newblock \bibinfo{title}{Fault-{Tolerant} {Quantum} {Computation} with {Constant} {Overhead}} (\bibinfo{year}{2014}).
\newblock \urlprefix\url{http://arxiv.org/abs/1310.2984}.
\newblock \bibinfo{note}{ArXiv:1310.2984 [quant-ph]}.

\bibitem{breuckmann_quantum_2021}
\bibinfo{author}{Breuckmann, N.~P.} \& \bibinfo{author}{Eberhardt, J.~N.}
\newblock \bibinfo{title}{Quantum {Low}-{Density} {Parity}-{Check} {Codes}}.
\newblock \emph{\bibinfo{journal}{PRX Quantum}} \textbf{\bibinfo{volume}{2}}, \bibinfo{pages}{040101} (\bibinfo{year}{2021}).
\newblock \urlprefix\url{https://link.aps.org/doi/10.1103/PRXQuantum.2.040101}.

\bibitem{bravyi_high-threshold_2024}
\bibinfo{author}{Bravyi, S.} \emph{et~al.}
\newblock \bibinfo{title}{High-threshold and low-overhead fault-tolerant quantum memory}.
\newblock \emph{\bibinfo{journal}{Nature}} \textbf{\bibinfo{volume}{627}}, \bibinfo{pages}{778--782} (\bibinfo{year}{2024}).
\newblock \urlprefix\url{http://arxiv.org/abs/2308.07915}.
\newblock \bibinfo{note}{ArXiv:2308.07915 [quant-ph]}.

\bibitem{wang_demonstration_2025}
\bibinfo{author}{Wang, K.} \emph{et~al.}
\newblock \bibinfo{title}{Demonstration of low-overhead quantum error correction codes} (\bibinfo{year}{2025}).
\newblock \urlprefix\url{http://arxiv.org/abs/2505.09684}.
\newblock \bibinfo{note}{ArXiv:2505.09684 [quant-ph]}.

\bibitem{andersen_small_2025}
\bibinfo{author}{Andersen, C.~K.} \& \bibinfo{author}{Greplová, E.}
\newblock \bibinfo{title}{Small {Quantum} {Low} {Parity} {Density} {Check} {Codes} for {Near}-{Term} {Experiments}} (\bibinfo{year}{2025}).
\newblock \urlprefix\url{http://arxiv.org/abs/2507.09690}.
\newblock \bibinfo{note}{ArXiv:2507.09690 [quant-ph]}.

\bibitem{voss_multivariate_2025}
\bibinfo{author}{Voss, L.}, \bibinfo{author}{Xian, S.~J.}, \bibinfo{author}{Haug, T.} \& \bibinfo{author}{Bharti, K.}
\newblock \bibinfo{title}{Multivariate bicycle codes}.
\newblock \emph{\bibinfo{journal}{Physical Review A}} \textbf{\bibinfo{volume}{111}}, \bibinfo{pages}{L060401} (\bibinfo{year}{2025}).
\newblock \urlprefix\url{https://link.aps.org/doi/10.1103/ll5p-z88p}.

\bibitem{liang_generalized_2025}
\bibinfo{author}{Liang, Z.}, \bibinfo{author}{Liu, K.}, \bibinfo{author}{Song, H.} \& \bibinfo{author}{Chen, Y.-A.}
\newblock \bibinfo{title}{Generalized toric codes on twisted tori for quantum error correction}.
\newblock \emph{\bibinfo{journal}{PRX Quantum}} \textbf{\bibinfo{volume}{6}}, \bibinfo{pages}{020357} (\bibinfo{year}{2025}).
\newblock \urlprefix\url{http://arxiv.org/abs/2503.03827}.
\newblock \bibinfo{note}{ArXiv:2503.03827 [quant-ph]}.

\bibitem{gong_toward_2024}
\bibinfo{author}{Gong, A.}, \bibinfo{author}{Cammerer, S.} \& \bibinfo{author}{Renes, J.~M.}
\newblock \bibinfo{title}{Toward {Low}-latency {Iterative} {Decoding} of {QLDPC} {Codes} {Under} {Circuit}-{Level} {Noise}} (\bibinfo{year}{2024}).
\newblock \urlprefix\url{http://arxiv.org/abs/2403.18901}.
\newblock \bibinfo{note}{ArXiv:2403.18901 [quant-ph]}.

\bibitem{muller_improved_2025}
\bibinfo{author}{Müller, T.} \emph{et~al.}
\newblock \bibinfo{title}{Improved belief propagation is sufficient for real-time decoding of quantum memory} (\bibinfo{year}{2025}).
\newblock \urlprefix\url{http://arxiv.org/abs/2506.01779}.
\newblock \bibinfo{note}{ArXiv:2506.01779 [quant-ph]}.

\bibitem{maurer_real-time_2025}
\bibinfo{author}{Maurer, T.} \emph{et~al.}
\newblock \bibinfo{title}{Real-time decoding of the gross code memory with {FPGAs}} (\bibinfo{year}{2025}).
\newblock \urlprefix\url{http://arxiv.org/abs/2510.21600}.
\newblock \bibinfo{note}{ArXiv:2510.21600 [quant-ph]}.

\bibitem{wolanski_ambiguity_nodate}
\bibinfo{author}{Wolanski, S.} \& \bibinfo{author}{Barber, B.}
\newblock \bibinfo{title}{Ambiguity {Clustering}: an accurate and efficient decoder for {qLDPC} codes} .

\bibitem{beni_tesseract_2025}
\bibinfo{author}{Beni, L.~A.}, \bibinfo{author}{Higgott, O.} \& \bibinfo{author}{Shutty, N.}
\newblock \bibinfo{title}{Tesseract: {A} {Search}-{Based} {Decoder} for {Quantum} {Error} {Correction}} (\bibinfo{year}{2025}).
\newblock \urlprefix\url{http://arxiv.org/abs/2503.10988}.
\newblock \bibinfo{note}{ArXiv:2503.10988 [quant-ph]}.

\bibitem{hillmann_localized_2025}
\bibinfo{author}{Hillmann, T.} \emph{et~al.}
\newblock \bibinfo{title}{Localized statistics decoding for quantum low-density parity-check codes}.
\newblock \emph{\bibinfo{journal}{Nature Communications}} \textbf{\bibinfo{volume}{16}}, \bibinfo{pages}{8214} (\bibinfo{year}{2025}).
\newblock \urlprefix\url{http://arxiv.org/abs/2406.18655}.
\newblock \bibinfo{note}{ArXiv:2406.18655 [quant-ph]}.

\bibitem{cohen_low-overhead_2022}
\bibinfo{author}{Cohen, L.~Z.}, \bibinfo{author}{Kim, I.~H.}, \bibinfo{author}{Bartlett, S.~D.} \& \bibinfo{author}{Brown, B.~J.}
\newblock \bibinfo{title}{Low-overhead fault-tolerant quantum computing using long-range connectivity}.
\newblock \emph{\bibinfo{journal}{Science Advances}} \textbf{\bibinfo{volume}{8}}, \bibinfo{pages}{eabn1717} (\bibinfo{year}{2022}).
\newblock \urlprefix\url{https://www.science.org/doi/10.1126/sciadv.abn1717}.

\bibitem{cross_improved_2025}
\bibinfo{author}{Cross, A.~W.}, \bibinfo{author}{He, Z.}, \bibinfo{author}{Rall, P.~J.} \& \bibinfo{author}{Yoder, T.~J.}
\newblock \bibinfo{title}{Improved {QLDPC} {Surgery}: {Logical} {Measurements} and {Bridging} {Codes}} (\bibinfo{year}{2025}).
\newblock \urlprefix\url{http://arxiv.org/abs/2407.18393}.
\newblock \bibinfo{note}{ArXiv:2407.18393 [quant-ph]}.

\bibitem{he_extractors_2025}
\bibinfo{author}{He, Z.}, \bibinfo{author}{Cowtan, A.}, \bibinfo{author}{Williamson, D.~J.} \& \bibinfo{author}{Yoder, T.~J.}
\newblock \bibinfo{title}{Extractors: {QLDPC} {Architectures} for {Efficient} {Pauli}-{Based} {Computation}} (\bibinfo{year}{2025}).
\newblock \urlprefix\url{http://arxiv.org/abs/2503.10390}.
\newblock \bibinfo{note}{ArXiv:2503.10390 [quant-ph]}.

\bibitem{yoder_tour_2025}
\bibinfo{author}{Yoder, T.~J.} \emph{et~al.}
\newblock \bibinfo{title}{Tour de gross: {A} modular quantum computer based on bivariate bicycle codes} (\bibinfo{year}{2025}).
\newblock \urlprefix\url{http://arxiv.org/abs/2506.03094}.
\newblock \bibinfo{note}{ArXiv:2506.03094 [quant-ph]}.

\bibitem{grassl_leveraging_2013}
\bibinfo{author}{Grassl, M.} \& \bibinfo{author}{Roetteler, M.}
\newblock \bibinfo{title}{Leveraging automorphisms of quantum codes for fault-tolerant quantum computation}.
\newblock In \emph{\bibinfo{booktitle}{2013 {IEEE} {International} {Symposium} on {Information} {Theory}}}, \bibinfo{pages}{534--538} (\bibinfo{publisher}{IEEE}, \bibinfo{address}{Istanbul, Turkey}, \bibinfo{year}{2013}).
\newblock \urlprefix\url{http://ieeexplore.ieee.org/document/6620283/}.

\bibitem{mcewen_relaxing_2023}
\bibinfo{author}{McEwen, M.}, \bibinfo{author}{Bacon, D.} \& \bibinfo{author}{Gidney, C.}
\newblock \bibinfo{title}{Relaxing {Hardware} {Requirements} for {Surface} {Code} {Circuits} using {Time}-dynamics}.
\newblock \emph{\bibinfo{journal}{Quantum}} \textbf{\bibinfo{volume}{7}}, \bibinfo{pages}{1172} (\bibinfo{year}{2023}).
\newblock \urlprefix\url{http://arxiv.org/abs/2302.02192}.
\newblock \bibinfo{note}{ArXiv:2302.02192 [quant-ph]}.

\bibitem{eickbusch_demonstration_2025}
\bibinfo{author}{Eickbusch, A.} \emph{et~al.}
\newblock \bibinfo{title}{Demonstration of dynamic surface codes}.
\newblock \emph{\bibinfo{journal}{Nature Physics}}  (\bibinfo{year}{2025}).
\newblock \urlprefix\url{https://www.nature.com/articles/s41567-025-03070-w}.

\bibitem{geher_error-corrected_2024}
\bibinfo{author}{Gehér, G.~P.}, \bibinfo{author}{McLauchlan, C.}, \bibinfo{author}{Campbell, E.~T.}, \bibinfo{author}{Moylett, A.~E.} \& \bibinfo{author}{Crawford, O.}
\newblock \bibinfo{title}{Error-corrected {Hadamard} gate simulated at the circuit level}.
\newblock \emph{\bibinfo{journal}{Quantum}} \textbf{\bibinfo{volume}{8}}, \bibinfo{pages}{1394} (\bibinfo{year}{2024}).
\newblock \urlprefix\url{http://arxiv.org/abs/2312.11605}.
\newblock \bibinfo{note}{ArXiv:2312.11605 [quant-ph]}.

\bibitem{gidney_inplace_2024}
\bibinfo{author}{Gidney, C.}
\newblock \bibinfo{title}{Inplace {Access} to the {Surface} {Code} {Y} {Basis}}.
\newblock \emph{\bibinfo{journal}{Quantum}} \textbf{\bibinfo{volume}{8}}, \bibinfo{pages}{1310} (\bibinfo{year}{2024}).
\newblock \urlprefix\url{http://arxiv.org/abs/2302.07395}.
\newblock \bibinfo{note}{ArXiv:2302.07395 [quant-ph]}.

\bibitem{shaw_lowering_2025}
\bibinfo{author}{Shaw, M.~H.} \& \bibinfo{author}{Terhal, B.~M.}
\newblock \bibinfo{title}{Lowering {Connectivity} {Requirements} {For} {Bivariate} {Bicycle} {Codes} {Using} {Morphing} {Circuits}}.
\newblock \emph{\bibinfo{journal}{Physical Review Letters}} \textbf{\bibinfo{volume}{134}}, \bibinfo{pages}{090602} (\bibinfo{year}{2025}).
\newblock \urlprefix\url{http://arxiv.org/abs/2407.16336}.
\newblock \bibinfo{note}{ArXiv:2407.16336 [quant-ph]}.

\bibitem{kovalev_quantum_2013}
\bibinfo{author}{Kovalev, A.~A.} \& \bibinfo{author}{Pryadko, L.~P.}
\newblock \bibinfo{title}{Quantum {Kronecker} sum-product low-density parity-check codes with finite rate}.
\newblock \emph{\bibinfo{journal}{Physical Review A}} \textbf{\bibinfo{volume}{88}}, \bibinfo{pages}{012311} (\bibinfo{year}{2013}).
\newblock \urlprefix\url{https://link.aps.org/doi/10.1103/PhysRevA.88.012311}.

\bibitem{lin_quantum_2024}
\bibinfo{author}{Lin, H.-K.} \& \bibinfo{author}{Pryadko, L.~P.}
\newblock \bibinfo{title}{Quantum two-block group algebra codes}.
\newblock \emph{\bibinfo{journal}{Physical Review A}} \textbf{\bibinfo{volume}{109}}, \bibinfo{pages}{022407} (\bibinfo{year}{2024}).
\newblock \urlprefix\url{https://link.aps.org/doi/10.1103/PhysRevA.109.022407}.

\bibitem{gidney_fault-tolerant_2021}
\bibinfo{author}{Gidney, C.}, \bibinfo{author}{Newman, M.}, \bibinfo{author}{Fowler, A.} \& \bibinfo{author}{Broughton, M.}
\newblock \bibinfo{title}{A {Fault}-{Tolerant} {Honeycomb} {Memory}}.
\newblock \emph{\bibinfo{journal}{Quantum}} \textbf{\bibinfo{volume}{5}}, \bibinfo{pages}{605} (\bibinfo{year}{2021}).
\newblock \urlprefix\url{http://arxiv.org/abs/2108.10457}.
\newblock \bibinfo{note}{ArXiv:2108.10457 [quant-ph]}.

\bibitem{beverland_fail_2025}
\bibinfo{author}{Beverland, M.~E.}, \bibinfo{author}{Carroll, M.}, \bibinfo{author}{Cross, A.~W.} \& \bibinfo{author}{Yoder, T.~J.}
\newblock \bibinfo{title}{Fail fast: techniques to probe rare events in quantum error correction} (\bibinfo{year}{2025}).
\newblock \urlprefix\url{http://arxiv.org/abs/2511.15177}.
\newblock \bibinfo{note}{ArXiv:2511.15177 [quant-ph]}.

\bibitem{roffe_decoding_2020}
\bibinfo{author}{Roffe, J.}, \bibinfo{author}{White, D.~R.}, \bibinfo{author}{Burton, S.} \& \bibinfo{author}{Campbell, E.}
\newblock \bibinfo{title}{Decoding across the quantum low-density parity-check code landscape}.
\newblock \emph{\bibinfo{journal}{Physical Review Research}} \textbf{\bibinfo{volume}{2}}, \bibinfo{pages}{043423} (\bibinfo{year}{2020}).
\newblock \urlprefix\url{https://link.aps.org/doi/10.1103/PhysRevResearch.2.043423}.

\bibitem{panteleev_degenerate_2021}
\bibinfo{author}{Panteleev, P.} \& \bibinfo{author}{Kalachev, G.}
\newblock \bibinfo{title}{Degenerate {Quantum} {LDPC} {Codes} {With} {Good} {Finite} {Length} {Performance}}.
\newblock \emph{\bibinfo{journal}{Quantum}} \textbf{\bibinfo{volume}{5}}, \bibinfo{pages}{585} (\bibinfo{year}{2021}).
\newblock \urlprefix\url{http://arxiv.org/abs/1904.02703}.
\newblock \bibinfo{note}{ArXiv:1904.02703 [quant-ph]}.

\bibitem{bravyi_tradeoffs_2010}
\bibinfo{author}{Bravyi, S.}, \bibinfo{author}{Poulin, D.} \& \bibinfo{author}{Terhal, B.}
\newblock \bibinfo{title}{Tradeoffs for {Reliable} {Quantum} {Information} {Storage} in {2D} {Systems}}.
\newblock \emph{\bibinfo{journal}{Physical Review Letters}} \textbf{\bibinfo{volume}{104}}, \bibinfo{pages}{050503} (\bibinfo{year}{2010}).
\newblock \urlprefix\url{https://link.aps.org/doi/10.1103/PhysRevLett.104.050503}.

\bibitem{tremblay_constant-overhead_2022}
\bibinfo{author}{Tremblay, M.~A.}, \bibinfo{author}{Delfosse, N.} \& \bibinfo{author}{Beverland, M.~E.}
\newblock \bibinfo{title}{Constant-{Overhead} {Quantum} {Error} {Correction} with {Thin} {Planar} {Connectivity}}.
\newblock \emph{\bibinfo{journal}{Physical Review Letters}} \textbf{\bibinfo{volume}{129}}, \bibinfo{pages}{050504} (\bibinfo{year}{2022}).
\newblock \urlprefix\url{https://link.aps.org/doi/10.1103/PhysRevLett.129.050504}.

\bibitem{gidney_stim_2021}
\bibinfo{author}{Gidney, C.}
\newblock \bibinfo{title}{Stim: a fast stabilizer circuit simulator}.
\newblock \emph{\bibinfo{journal}{Quantum}} \textbf{\bibinfo{volume}{5}}, \bibinfo{pages}{497} (\bibinfo{year}{2021}).
\newblock \urlprefix\url{http://arxiv.org/abs/2103.02202}.
\newblock \bibinfo{note}{ArXiv:2103.02202 [quant-ph]}.

\bibitem{javadi-abhari_big_2025}
\bibinfo{author}{Javadi-Abhari, A.}, \bibinfo{author}{Martiel, S.}, \bibinfo{author}{Seif, A.}, \bibinfo{author}{Takita, M.} \& \bibinfo{author}{Wei, K.~X.}
\newblock \bibinfo{title}{Big cats: entanglement in 120 qubits and beyond} (\bibinfo{year}{2025}).
\newblock \urlprefix\url{http://arxiv.org/abs/2510.09520}.
\newblock \bibinfo{note}{ArXiv:2510.09520 [quant-ph]}.

\bibitem{bravyi_simulation_2013}
\bibinfo{author}{Bravyi, S.} \& \bibinfo{author}{Vargo, A.}
\newblock \bibinfo{title}{Simulation of rare events in quantum error correction}.
\newblock \emph{\bibinfo{journal}{Physical Review A}} \textbf{\bibinfo{volume}{88}}, \bibinfo{pages}{062308} (\bibinfo{year}{2013}).
\newblock \urlprefix\url{http://arxiv.org/abs/1308.6270}.
\newblock \bibinfo{note}{ArXiv:1308.6270 [quant-ph]}.

\end{thebibliography}
\end{appendix}

\end{document}